\documentclass{article}

\usepackage{arxiv}
\usepackage{float}
\usepackage[utf8]{inputenc} 
\usepackage[T1]{fontenc}    
\usepackage{hyperref}       
\usepackage{url}            
\usepackage{booktabs}       
\usepackage{amsfonts}       
\usepackage{nicefrac}       
\usepackage{microtype}      
\usepackage{xcolor}         
\usepackage{lipsum}
\usepackage{graphicx}
\usepackage{amsmath}
\usepackage{amssymb}
\usepackage{physics}
\usepackage{geometry}
\usepackage{graphicx}
\usepackage{algorithm}
\usepackage{algpseudocode}
\usepackage{subcaption}
\graphicspath{ {./images/} }

\title{Differentiable fast far-field transform in cylindrical coordinates for large-area cascaded metalens optics}

\author{
Arvin Keshvari$^{1}$\thanks{Corresponding author: arvink@vt.edu},
Ata Shakeri$^{1}$,
William Tuxbury$^{1}$,
Joon-Suh Park$^{2}$,
Qing Wang$^{2}$,
Wei-Ting Chen$^{2}$,
Zin Lin$^{1}$\\
\\
$^{1}$Bradley Department of Electrical and Computer Engineering, Virginia Tech, Blacksburg, VA 24061, USA \\
$^{2}$SNOChip Inc., Princeton, NJ 08540, USA
}

\date{}

\begin{document}
\maketitle
\begin{abstract}
We present a fully differentiable far-field transform in cylindrical coordinates for efficient full-area point spread function (PSF) evaluation and optimization of large axisymmetric metalenses. The method computes wave-optical responses of apertures spanning thousands to tens of thousands of wavelengths in diameter---reaching millimeter scales in the visible and centimeter scales in the infrared---in seconds, achieving three to four orders of magnitude speedup over Green's function integration while avoiding the prohibitive memory of two-dimensional fast Fourier transforms. The approach decomposes vectorial near fields into angular-momentum channels processed in parallel, applies FFTLog-accelerated Hankel transforms, and employs Graf's addition theorem to recenter focal fields under oblique illumination. Analytic gradients are derived throughout using the adjoint method, enabling optimization with only ${\sim}65\%$ overhead relative to a forward simulation. For a 4\,mm-diameter aperture (${\sim}8000\lambda$, ${\sim}12{,}600$ azimuthal modes) at $30^\circ$ incidence, a forward--adjoint iteration requires only ${\sim}12$\,s on a 350-thread CPU, making large-scale oblique wave-optics optimization practical without ray-tracing approximations. Applied to polychromatic RGB (446/530/650\,nm) metalens design at normal incidence, full-area PSF evaluation reveals efficiency limitations obscured by conventional cropped focal-spot analyses: a mono-pillar metalens that appears diffraction-limited achieves only ${\sim}6\%$ average absolute focusing efficiency, whereas direct far-field optimization raises this to $37\%$ (locally periodic approximation) and $51\%$ (zoned discrete axisymmetry). A cascaded double-metasurface design reaches $63\%$, while a four-metasurface architecture attains $96\%$ average relative efficiency. We further demonstrate millimeter-scale oblique-incidence optimization of single-surface and doublet architectures, where cascaded doublets enable partial coma correction inaccessible to a single rotationally symmetric surface.
\end{abstract}


\section{Introduction}
State-of-the-art designs of compound metalenses are dominantly based on ray-tracing software (e.g., Zemax) to model light propagation across successive metalens layers and into the focal plane~\cite{Groever2017}. While ray tracing is computationally efficient, it neglects essential diffraction effects. On the other hand, wave-optics treatments, such as Huygens’ diffraction integrals, are prohibitively expensive for repeated evaluations within an iterative optimization loop. Black-box or legacy ray-tracing tools further limit seamless integration with diverse metasurface transmission models, including locally periodic surrogates~\cite{Pestourie2018}, domain-decomposition and full-wave simulations~\cite{sun2025scalable}. Recent work~\cite{Isnard:25} enables optimization through ray-traced, metasurface-refractive hybrids but still incurs residual approximations due to repeated back-and-forth transitions between geometric and wave optics.  

A particularly consequential limitation of existing simulation tools is their inability to efficiently compute \textit{full-area} point spread functions (PSFs).The PSF is the diffraction pattern evaluated across the entire focal plane, not just a cropped region around the central focal spot. Full-area PSFs are essential for end-to-end co-design of metaoptics and computational deconvolution~\cite{lin2022end}.  By offering  a complete measure of focusing quality, the full-area PSF enables a thorough assessment of stray light inefficiencies and the latent noise floor far from the focal spot. In practice, most metalens studies report only a cropped central region of the PSF, which can present a misleadingly sharp focal spot while concealing an elevated noise floor at larger radii. The ability to rapidly compute and assess full-area PSFs is therefore not merely a nice-to-have computational option but an imperative for meaningful optimization of cutting-edge metalens designs.

To accelerate far-field propagation Fast Fourier Transforms (FFTs) in Cartesian coordinates can be used, exploiting the convolution structure of the Green’s function. Indeed, FFT-based propagation is routinely used in computational imaging models for diffractive optical elements and phase masks~\cite{Goodman2017}. However, for large-area metalenses with millimeter-to-centimeter-scale apertures and sub-wavelength sampling pitch, two-dimensional FFTs become infeasible: resolving a 1~cm aperture at a pitch of $\lambda/5$ requires arrays of $\sim 10^5 \times 10^5$ elements, consuming over a terabyte of memory for a single complex field. Moreover, metalenses are structurally axisymmetric. Hence their transmitted fields can be fully characterized by one-dimensional radial functions. Using two-dimensional transforms to process intrinsically one-dimensional data is fundamentally wasteful.

In this paper, we present a fully differentiable, fast far-field transform in cylindrical coordinates $(r,\theta,z)$ for simulation and optimization of light propagation through axisymmetric metalenses, eliminating the need for ray tracing entirely in metalens design pipeline. Our approach is based on vector cylindrical harmonics (VCHs)~\cite{Stratton1941}, which provide a natural basis for wave propagation in axisymmetric structures. VCHs decompose two-dimensional Cartesian fields into one-dimensional radial functions over a finite set of discrete angular momentum channels $m\in \{-M_{\text{max}}, ..., M_\text{max}\}$, each of which can be processed independently and in parallel. Just as the far-field transform in the plane-wave basis employs a two-dimensional Fourier transform, the far-field transform in the VCH basis employs one-dimensional Hankel transforms~\cite{beckman2024nonuniform}, one per angular momentum channel. For normal incidence on an axisymmetric metalens, only two channels ($m = \pm 1$) are required, reducing the computational cost by a factor of $\sim N$ (the number of radial grid points) relative to a 2D FFT. Even at large oblique angles, where thousands of $m$-modes contribute, the dimensional reduction yields orders-of-magnitude savings in both memory and computation. We further accelerate each Hankel transform using an FFTLog algorithm~\cite{2015ascl.soft12017H}, an $O(N \log N)$ method based on a logarithmically spaced radial grid. We implemented the entire pipeline, including custom parallel vector--Jacobian products, in Julia with shared-memory multithreading across angular momentum channels.

A key contribution of this work is that our cylindrical far-field transform is \textit{fully differentiable}: we derive the adjoint of each stage of the pipeline and implement parallel reverse-mode gradient propagation, enabling scalable optimization of general loss functionals defined on the PSF. The adjoint adds only $\sim$65\% overhead to the forward evaluation, so that a complete forward--adjoint iteration through a 4~mm-diameter metalens at $30^\circ$ oblique incidence ($\sim$12{,}600 angular momentum modes) executes in approximately 12~seconds on a 350-thread CPU. This speed makes gradient-based wave-optic optimization of millimeter-scale metalens PSFs, including cascaded multi-surface architectures,  routine within a few hours rather than over several weeks.

We illustrate the practical payoff of this capability in two ways. First, full-area PSF evaluation lets us audit and improve focusing efficiency across a hierarchy of designs (Fig.~\ref{fig:Fig5_new}): a hand-designed, ray-traced mono-pillar RGB metalens achieves only ${\sim}6\%$ average absolute focusing efficiency when its true full-area PSF is computed, whereas directly optimizing the far field with our differentiable transform raises this to $37.11\%$ (locally periodic approximation) and $50.93\%$ (zoned discrete axisymmetry), and a cascaded double-metasurface design reaches $62.66\%$. Second, the speed of the pipeline makes deep cascades tractable: a four-metasurface RGB metalens optimized at normal incidence attains $96\%$ average relative focusing efficiency (Fig.~\ref{fig:Fig45_new}), with each three-wavelength forward PSF evaluation over $47{,}630$ radial grid points taking only $1.043$~seconds (Table~\ref{tab:timings}). To the best of our knowledge, this is the first wave-optics optimization of a four-metasurface, large-area (several-millimeter-diameter) metalens performed without any ray-tracing approximation.

While the principle of far-field propagation in the basis of vector cylindrical harmonics is not new, the relevant results are scattered throughout decades-old literature~\cite{Stratton1941,Harrington1961}. To the best of our knowledge, the method has never been accelerated using a dedicated Fast Hankel Transform, nor exploited for \textit{optimization} (not just simulation) of large-scale apertures. Our work consolidates these results into a differentiable computational package that can be seamlessly integrated with metasurface transmission models and end-to-end optimization pipelines. This framework enables us to demonstrate the first wave-optics optimization of cascaded metasurfaces with aperture diameters of several millimeters under oblique incidence . 

The remainder of this paper is organized as follows: In Sec.~\ref{sec:cyFFT}, we present the theoretical framework and the computational techniques that underpin our fast cylindrical far-field transform. Sec.~\ref{sec:3} demonstrates polychromatic metalens design at normal incidence, including single-surface and cascaded architectures, and validates the pipeline under oblique illumination using proxy phase profiles. Finally, Sec.~\ref{sec:conclusion} concludes with an outlook on integrating angle-dependent metasurface models.

\section{Theoretical and computational methods}
\label{sec:cyFFT}
\subsection{Far field transform in cylindrical coordinates}

We consider propagating an electric field from the near plane at $z=0$ to an arbitrary focal plane at distance $z=f$, through a uniform medium of refractive index $n$. The following formulation is an exact consequence of Maxwell's equations in a source-free medium---no paraxial, Fresnel, or thin-lens approximation is invoked in the far-field propagation. The vectorial near-field can be expanded in the basis of vector cylindrical harmonics (VCHs), whose expressions are derived in Appendix~\ref{apx:vch}~\cite{Stratton1941}:
\begin{align}
\mathbf{E}_{\text{Near}}(r, \theta) =
\mathbf{E}(r, \theta, z{=}0) &=
\sum_m \mathbf{E}_m^\text{Near}(r)\, e^{im\theta}  \notag \\
&=\sum_{m}\int dk_{r} \left[ A_{m}^{\text{TE}}(k_{r})\, \mathbf{E}_{m,k_{r}}^{\text{TE}}(r)\, e^{i m \theta}+ A_{m}^{\text{TM}}(k_{r})\, \mathbf{E}_{m,k_{r}}^{\text{TM}}(r)\, e^{i m \theta} \right]. \label{eq:Enear}
\end{align}
Here, TE (TM) denotes transverse electric (transverse magnetic) modes. Each harmonic is indexed by a discrete azimuthal order $m \in \{-M_{\text{max}},\ldots,+M_{\text{max}}\}$ (corresponding to the orbital angular momentum harmonic $e^{i m \theta}$) and a continuous radial wavenumber $k_r$. The expansion coefficients are obtained by projecting the near field onto the VCH basis (see Appendix~\ref{apx:coefs}):
\begin{align}
    A^\text{TE}_m(k_r) &= \frac{1}{k_r} \int_0^\infty \mathbf{E}_m^\text{Near}(r) \cdot \mathbf{E}_{m,k_{r}}^{\text{TE}*}(r) \; r \, dr, \label{eq:ATE} \\
    A^\text{TM}_m(k_r) &= \frac{\omega^2 \varepsilon^2}{k_z^2 k_r} \int_0^\infty \mathbf{E}_m^\text{Near}(r) \cdot \mathbf{E}_{m,k_{r}}^{\text{TM}*}(r) \; r \, dr. \label{eq:ATM}
\end{align}
Substituting the explicit VCH basis functions (Eqs.~\ref{eq:ETM}--\ref{eq:ETE} in Appendix~\ref{apx:vch}) into these projections, the dot products with $\hat{r}$ and $\hat{\theta}$ components reduce each integral to terms of the form $\int_0^\infty f_m(r)\,J_m(k_r r)\,r\,dr$, i.e., Hankel transforms of order $m$. This identification is what enables the use of fast Hankel transform algorithms in the next sub-section.

The far field is then given by
\begin{align}
\mathbf{E}_{\text{Far}}(r, \theta, z) &=\sum_{m}\int dk_{r} \left[ A_{m}^{\text{TE}}(k_{r})\, \mathbf{E}_{m,k_{r}}^{\text{TE}}(r)\, e^{i m \theta}\, e^{i k_z z}+ A_{m}^{\text{TM}}(k_{r})\, \mathbf{E}_{m,k_{r}}^{\text{TM}}(r)\, e^{i m \theta}\, e^{i k_z z} \right] \label{eq:kz}
\end{align}
where \( k_z(k_r) = \sqrt{k^2 - k_r^2} \) and \( k = \frac{2 \pi n}{\lambda}\) are wavevector magnitude and component along the optical axis, respectively.
\subsection{Computational pipeline and acceleration techniques}
\label{sec:tricks}
Based on the preceding analytical formulation, we now describe the computational pipeline that evaluates the far field and the techniques employed to accelerate the calculation. The complete forward algorithms for single-surface and cascaded (doublet) pipelines are summarized in Algorithms~\ref{alg:fwd_single} and~\ref{alg:fwd_doublet}, respectively; their adjoint counterparts are given in Algorithms~\ref{alg:adj_single} and~\ref{alg:adj_doublet}.

\paragraph{Step 1: Azimuthal decomposition.}
We first decompose the near field into azimuthal harmonics $e^{im\theta}$ to obtain the radial components $\mathbf{E}_m^\text{Near}(r)$. For a plane wave incident on a \textit{local}, axisymmetric metalens surface, the azimuthal decomposition of the transmitted near field can be carried out analytically via the Jacobi--Anger expansion (see Appendix~\ref{apx:pwe}); otherwise, it can be efficiently obtained via a one-dimensional Fourier transform in $\theta$.

\paragraph{Step 2: Forward Hankel transform.}
The spectral coefficients $A^\text{TE,TM}_m(k_r)$ are evaluated through the projection integrals (Eqs.~\ref{eq:ATE}--\ref{eq:ATM}), which reduce to Hankel transforms of order $m$ (Sec.~\ref{sec:cyFFT}). We evaluate each Hankel transform using the FFTLog algorithm~\cite{2015ascl.soft12017H}, an $O(N\log N)$ method that operates on logarithmically spaced radial grids. Log-spacing in both $r$ and $k_r$ turns the Hankel transform kernel $J_m(k_r r)$ into a convolution in log-space, which can then be evaluated via a standard FFT. As a secondary benefit, the log-spaced grid naturally provides fine resolution near the optical axis and coarse resolution at the aperture edge.

\paragraph{Step 3: Propagation.}
Each spectral coefficient is multiplied by the propagation phase $e^{ik_z(k_r)\,f}$ (Eq.~\ref{eq:kz}), advancing the field to the focal plane. This is a pointwise, $O(N)$ operation per mode.

\paragraph{Step 4: Focal-plane re-centering via Graf’s addition theorem.}
For normal incidence, the focal spot lies at the origin, and no re-centering is needed: the inverse Hankel transform (Step~5) directly yields the PSF on the original radial grid, covering the \textit{full} focal plane. For oblique incidence, however, the focal spot is displaced to $(f\tan\alpha, 0)$. Evaluating the field near this shifted focus using the origin-centered harmonics would require synthesizing the entire focal plane. Instead, we exploit the non-uniform sampling of the FFTLog to concentrate resolution on the shifted region. Graf’s addition theorem~\cite{lozier2003nist} overcomes this by re-expanding the propagated harmonics in a \textit{local} cylindrical coordinate system $(\rho,\psi)$ centered at the focal spot (see Appendix~\ref{apx:graf} for the full derivation). The re-expansion takes the form of a mode convolution,
\begin{equation}
    B_l(k_r) = \sum_{m} \tilde{A}_{m+l}(k_r)\, J_m(k_r x_0), \label{eq:graf}
\end{equation}
where $x_0 = f\tan\alpha$ is the focal shift, $J_m(k_r x_0)$ are Bessel weights, and $\tilde{A}_m(k_r) = A_m(k_r)\,e^{ik_z f}$ are the propagated spectral coefficients. Only $L_{\text{max}} \ll M_\text{max}$ local modes are needed to resolve the PSF (e.g., $L_{\text{max}} \approx 15$ for a $5\lambda$ output region), and Bessel truncation at each $k_r$ limits the effective sum length, bounding the compute cost despite the large number of input modes. The local coefficients $B_l(k_r)$ feed into the inverse Hankel transform exactly as in the normal-incidence case, which does not require re-centering.

\paragraph{Step 5: Inverse Hankel transform and angular synthesis.}
The local spectral coefficients $B_l(k_r)$ (or $\tilde{A}_m(k_r)$ at normal incidence) are transformed back to spatial coordinates via inverse FFTLog calls, producing radial profiles $b_l(\rho)$ on the focal-plane grid. An inverse FFT over the local azimuthal index $l$ then reconstructs the two-dimensional field $u(\rho,\psi) = \sum_l b_l(\rho)\,e^{il\psi}$, from which the PSF intensity $|u|^2$ is obtained.

\paragraph{Extension to cascaded metasurfaces.}
The pipeline above describes propagation through a single surface to the focal plane. For cascaded architectures with two or more metasurfaces separated by a distance $d$, the pipeline is extended by inserting additional Hankel round-trips between surfaces: after Step~2, the spectral coefficients are propagated by $d$ (Step~3), transformed back to spatial coordinates via an inverse FFTLog, multiplied by the second surface's transmission $t_2(r)$, and then transformed forward again before continuing to the focal plane via Steps~3--5. In the cascaded case the final propagation distance is $f{-}d$, so the propagated coefficients entering Steps~4--5 are $\tilde{A}^{(2)}_m = A^{(2)}_m\,e^{ik_z(f-d)}$; throughout, the tilde denotes coefficients propagated from the last surface to the focal plane. Each inter-surface round-trip adds one forward and one inverse Hankel transform per $m$-mode, and the entire extension inherits the same per-mode parallelism as the single-surface pipeline.

\paragraph{Parallelism.}
In an axisymmetric metalens, the transmitted field of a normally incident plane wave contains only two angular momentum components ($m = \pm 1$). For oblique incidence, the number of contributing modes grows (e.g., $M_\text{max} \approx 12{,}600$ for a 4~mm aperture at $\alpha = 30^\circ$), but each $m$-mode can be processed independently through Steps~2--3. This embarrassingly parallel structure~\cite{Christiansen2021inverse} scales linearly with the number of threads or processors and requires no inter-process communication, in contrast to distributed 2D FFT approaches which require costly all-to-all transposes. A further $2\times$ speedup is obtained by exploiting the symmetry $\tilde{A}_{-m} = (-1)^m \tilde{A}_m$, which allows the pipeline to process only $m \geq 0$ and reconstruct the negative-$m$ coefficients analytically (see Appendix~\ref{apx:negm}).

\paragraph{Differentiability.}
Because our objective is not only to evaluate but also to \textit{optimize} arbitrary functionals of the PSF, the entire pipeline must be rapidly differentiable. We derive the adjoint (reverse-mode derivative) of each step and implement parallel vector--Jacobian products that mirror the forward pipeline’s threading structure (see Appendix~\ref{apx:backprop} and Algorithms~\ref{alg:adj_single}--\ref{alg:adj_doublet}). The adjoint adds only $\sim$65\% overhead to the forward evaluation, enabling gradient-based optimization at negligible additional cost per iteration.

\paragraph{Computational complexity.}
The total cost of the forward pipeline is dominated by $O(M_\text{max}\, N_r \log N_r)$ and $O(L_{\text{max}}\, N_r \log N_r)$ for the FFTLog calls (Steps~2 and~5, respectively) and $O(M_\text{max}L_{\text{max}}\, N_r)$ for the Graf convolution (Step~4). By comparison, a two-dimensional Cartesian FFT over an equivalent grid costs $O(N_r^2 \log N_r)$. Since $M_\text{max} \ll N_r$ in all practical metalens problems (e.g., $M_\text{max} \approx 12{,}600$ versus $N_r \approx 131{,}000$ for a 4~mm aperture at $30^\circ$), the cylindrical pipeline is faster by a factor of roughly $N_r / M_\text{max}$, with correspondingly reduced memory.

\begin{algorithm}[t]
\caption{Single-surface forward pipeline}\label{alg:fwd_single}
\begin{algorithmic}[1]
\Require transmission $t(r)$, plan (precomputed grids, kernels, Bessel matrix)
\Ensure PSF intensity $I(\rho,\psi)$
\Statex \textit{--- Step 1: Mode construction (parallel over $m = 0,\ldots,M_\text{max}$) ---}
\For{each $m$}
    \State $u_m(r) \gets i^m\, t(r)\, J_m(k_x r)$ \Comment{Jacobi-Anger; Eq.~\ref{eq:Em_near}}
\EndFor
\Statex \textit{--- Step 2: Forward Hankel transform (parallel over $m$) ---}
\For{each $m$}
    \State $A_m(k_r) \gets \mathcal{H}_m\!\left[r\,u_m\right](k_r) \;/\; k_r$ \Comment{FFTLog; Eq.~\ref{eq:bessel_ortho}}
\EndFor
\Statex \textit{--- Steps 3+4 fused: Propagation + Graf shift (parallel over $k_r$) ---}
\For{each $k_r$}
    \State $\tilde{A}_m \gets A_m \cdot e^{ik_z f}$ for all $m$; \quad $\tilde{A}_{-m} \gets (-1)^m \tilde{A}_m$ \Comment{Eq.~\ref{eq:kz}; Appendix~\ref{apx:negm}}
    \State $B_l(k_r) \gets \sum_{m} \tilde{A}_{m+l}(k_r)\, J_m(k_r x_0)$ for $l = -L_{\text{max}},\ldots,L_{\text{max}}$ \Comment{Graf; Eq.~\ref{eq:graf}}
\EndFor
\Statex \textit{--- Step 5: Inverse Hankel + angular synthesis (parallel over $l$) ---}
\For{each $l = -L_{\text{max}},\ldots,L_{\text{max}}$}
    \State $b_l(\rho) \gets \mathcal{H}_l^{-1}\!\left[k_r\,B_l\right](\rho) \;/\; \rho$ \Comment{inverse FFTLog}
\EndFor
\State $u(\rho,\psi) \gets \sum_l b_l(\rho)\,e^{il\psi}$ \Comment{inverse FFT over $l$}
\State $I(\rho,\psi) \gets |u(\rho,\psi)|^2$
\end{algorithmic}
\end{algorithm}

\begin{algorithm}[t]
\caption{Doublet forward pipeline (two surfaces separated by $d$)}\label{alg:fwd_doublet}
\begin{algorithmic}[1]
\Require transmissions $t_1(r)$, $t_2(r)$; inter-surface distance $d$; plan
\Ensure PSF intensity $I(\rho,\psi)$
\Statex \textit{--- Steps 1--2d fused: Inter-surface propagation (parallel over $m$) ---}
\For{each $m = 0,\ldots,M_\text{max}$}
    \State $u_m^{(1)}(r) \gets i^m\, t_1(r)\, J_m(k_x r)$ \Comment{mode construction, surface 1}
    \State $\text{raw}_1(k_r) \gets \mathcal{H}_m\!\left[r\,u_m^{(1)}\right](k_r)$ \Comment{forward FFTLog}
    \State $\text{raw}_1 \gets \text{raw}_1 \cdot e^{ik_z d}$ \Comment{propagate to surface 2}
    \State $\text{raw}_2(r) \gets \mathcal{H}_m^{-1}\!\left[\text{raw}_1\right](r)$ \Comment{inverse FFTLog}
    \State Save $u_\text{mid}(r) \gets \text{raw}_2 / r$ \Comment{field at surface 2 (for adjoint)}
    \State $A_m^{(2)}(k_r) \gets \mathcal{H}_m\!\left[t_2(r) \cdot \text{raw}_2\right](k_r) \;/\; k_r$ \Comment{forward FFTLog through $t_2$}
\EndFor
\Statex \textit{--- Steps 3--5: Propagation by $(f{-}d)$ + Graf shift + inverse Hankel + synthesis ---}
\State Same as Algorithm~\ref{alg:fwd_single} Steps~3+4 and~5, using $A_m^{(2)}$ and propagation distance $f{-}d$
\end{algorithmic}
\end{algorithm}

\begin{algorithm}[t]
\caption{Single-surface adjoint pipeline}\label{alg:adj_single}
\begin{algorithmic}[1]
\Require cotangent $\overline{I}(\rho,\psi) \equiv \partial\mathcal{L}/\partial I$; saved forward intermediates
\Ensure gradient $\partial\mathcal{L}/\partial \bar{t}$ (Wirtinger derivative w.r.t.\ $\bar{t}$)
\Statex \textit{--- Adjoint of Steps 5 + intensity ---}
\State $\bar{u} \gets \overline{I} \odot u$ \Comment{adjoint of $|u|^2$}
\State $\bar{b}_l(\rho) \gets \text{FFT}_\psi[\bar{u}]$ for each $l$ \Comment{adjoint of angular synthesis}
\Statex \textit{--- Adjoint of inverse Hankel (parallel over $l$) ---}
\For{each $l = -L_{\text{max}},\ldots,L_{\text{max}}$}
    \State $\bar{B}_l(k_r) \gets k_r \cdot \mathcal{H}_l^\dagger\!\left[\bar{b}_l / \rho\right](k_r)$ \Comment{adjoint FFTLog (conjugate kernel)}
\EndFor
\Statex \textit{--- Adjoint of Steps 3+4: Graf shift + propagation (parallel over $k_r$) ---}
\For{each $k_r$}
    \State $\bar{A}_n(k_r) \gets e^{-ik_zf}\sum_l \bar{B}_l(k_r)\, J_{n-l}(k_r x_0)$ \Comment{transposed Graf convolution + adjoint propagation}
    \State $\bar{A}_m \gets \tfrac{\epsilon_m}{2}\left(\bar{A}_m + (-1)^m\,\bar{A}_{-m}\right)$ for $m \geq 0$ \Comment{adjoint of $\tilde{A}_{-m} = (-1)^m\tilde{A}_m$; $\epsilon_0{=}1$, $\epsilon_{m\geq1}{=}2$}
\EndFor
\Statex \textit{--- Adjoint of Steps 1+2 fused: FFTLog + mode construction (parallel over $m$) ---}
\For{each $m = 0,\ldots,M_\text{max}$}
    \State $\bar{g}_m(r) \gets \mathcal{H}_m^\dagger\!\left[\bar{A}_m / k_r\right](r)$ \Comment{adjoint FFTLog}
    \State $\partial\mathcal{L}/\partial \bar{t} \mathrel{+}= \overline{i^m}\, J_m(k_x r)\, r\, \bar{g}_m(r)$ \Comment{accumulate gradient}
\EndFor
\end{algorithmic}
\end{algorithm}

\begin{algorithm}[t]
\caption{Doublet adjoint pipeline}\label{alg:adj_doublet}
\begin{algorithmic}[1]
\Require cotangent $\overline{I}$; saved forward intermediates (including $u_\text{mid}$, $t_2$)
\Ensure gradients $\partial\mathcal{L}/\partial \bar{t}_1$, $\partial\mathcal{L}/\partial \bar{t}_2$
\Statex \textit{--- Adjoint of Steps 5 + intensity + inverse Hankel ---}
\State Same as Algorithm~\ref{alg:adj_single} through $\bar{B}_l(k_r)$
\Statex \textit{--- Adjoint of Steps 3+4: Graf shift + propagation by $(f{-}d)$ ---}
\State Same as Algorithm~\ref{alg:adj_single}, producing $\bar{A}_m^{(2)}$ using distance $f{-}d$
\Statex \textit{--- Adjoint of Steps 1--2d fused (parallel over $m$) ---}
\For{each $m = 0,\ldots,M_\text{max}$}
    \State $\bar{g}(r) \gets \mathcal{H}_m^\dagger\!\left[\bar{A}_m^{(2)} / k_r\right](r)$   \Comment{adjoint of surface-2 FFTLog}
    \State $\partial\mathcal{L}/\partial \bar{t}_2 \mathrel{+}= \overline{u_\text{mid}} \cdot r \cdot \bar{g}(r)$ \Comment{gradient w.r.t.\ $t_2$}
    \State $\bar{g}(r) \gets \bar{t}_2(r) \cdot \bar{g}(r)$ \Comment{propagate through $t_2$ multiplication}
    \State $\bar{g}(k_r) \gets \mathcal{H}_m^\dagger\!\left[\bar{g}\right](k_r) \cdot e^{-ik_z d}$ \Comment{adjoint inverse Hankel + propagation}
    \State $\bar{g}(r) \gets \mathcal{H}_m^\dagger\!\left[\bar{g}\right](r)$ \Comment{adjoint of surface-1 FFTLog}
    \State $\partial\mathcal{L}/\partial \bar{t}_1 \mathrel{+}= \overline{i^m}\, J_m(k_x r)\, r\, \bar{g}(r)$ \Comment{gradient w.r.t.\ $t_1$}
\EndFor
\end{algorithmic}
\end{algorithm}

\subsection{Metasurface transmission model}
\label{sec:metamodel}
Our far-field transform accepts an arbitrary near-field profile $\mathbf{E}_\text{Near}(r,\theta)$ as input and is therefore agnostic to the metasurface model that generates it. In the most general locally periodic approximation~\cite{Isnard:25}, the transmitted field through a metasurface is
\begin{align}
    \mathbf{E}_\text{out}(r,\theta) &= t(r,\theta;\lambda)\, \mathbf{E}_\text{in}(r,\theta), \label{eq:LPA}\\
    t(r,\theta;\lambda) &= t\!\left(\mathbf{g}(r,\theta),\,\alpha_\text{loc}(r,\theta),\,\beta_\text{loc}(r,\theta);\,\lambda\right), \label{eq:t_full}
\end{align}
where $\mathbf{g}(r,\theta)$ are the geometrical parameters of the meta-atom at position $(r,\theta)$, and $\alpha_\text{loc}$, $\beta_\text{loc}$ are the polar and azimuthal components of the \textit{local} angle of incidence as experienced by that meta-atom. The local angles are not simply the global incidence angle $\alpha$: they vary across the aperture because each meta-atom intercepts the incident wavefront at a different position and orientation. For the first surface illuminated by a plane wave, the local angles can be inferred from the known wavevector; for deeper surfaces in a cascaded stack, the incident field is the far field of the preceding surface, and the local propagation direction must be extracted from the phase gradient of this field via an Eikonal (ray-optics) approximation (see Appendix~\ref{apx:LPA} for the detailed construction). The transmission function for each meta-atom is obtained by simulating it in a periodic unit cell with Bloch boundary conditions set by $\alpha_\text{loc}$ and $\beta_\text{loc}$~\cite{Lalanne1998}. In principle, the discrete simulation data can be interpolated to build a surrogate model $t(\mathbf{g}, \alpha_\text{loc}, \beta_\text{loc};\lambda)$~\cite{PestourieSurrogate2022}. However, the full model (Eq.~\ref{eq:t_full}) is challenging to deploy at scale: accurately training surrogate functions over the joint space of geometry, two angular variables, and wavelength remains an open problem, particularly for the tall, high-aspect-ratio pillars required for full $2\pi$ phase coverage~\cite{Mansouree2021} (see also Appendix~\ref{apx:LPA}). Unlike shallow gratings, tall pillar meta-atoms support strongly resonant modes whose transmission varies non-monotonically with angle, making standard polynomial or neural-network interpolation unreliable over the required angular range. For this reason, the community has largely relied on simplified models. The most common simplification is to neglect angular dependence altogether, setting $t(\mathbf{g};\lambda)$~\cite{Pestourie2018}. This approximation is exact for normal incidence on an axisymmetric metalens, where every meta-atom sees the same incident angle, and it remains reasonable for first-surface illumination under moderate oblique angles. But, it breaks down for second or deeper surfaces in a cascaded stack, where the incident field is no longer a plane wave and the local angle varies across the aperture.
\begin{figure}[t]
  \centering
  \includegraphics[width=0.8\linewidth]{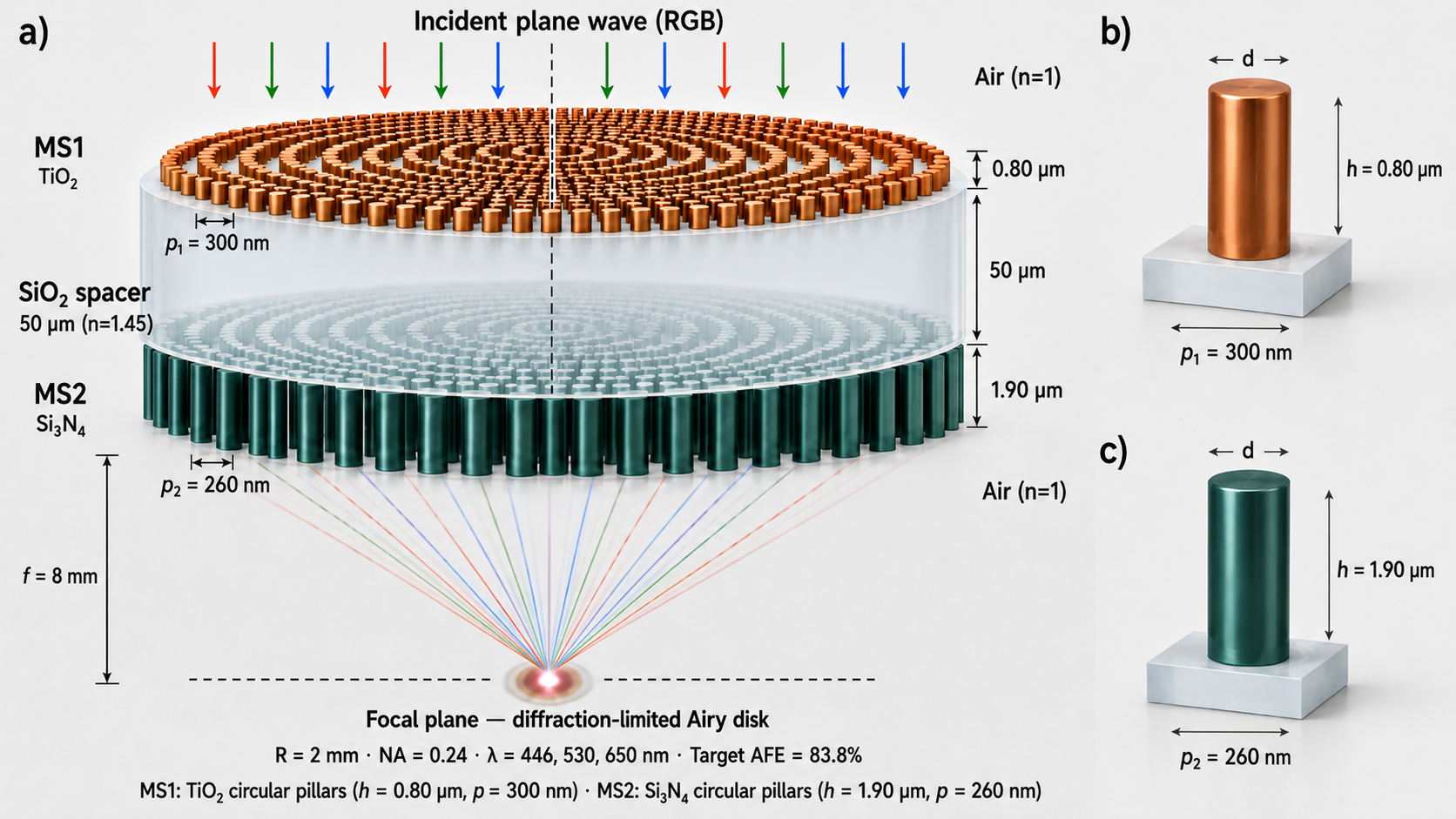}
  \caption{a) Axis-symmetric metalens structure. MS1 consists of 0.8 um long TiO2 pillars, then light propagates in SiO2 substrate to reach MS2, which consists of 1.9 um long Si3N4 pillars. Finally it propagates in air to reach the focal plane. b-c) Unit-cell for MS1 and MS2 structures, respectively.}
  \label{fig:Fig1}
\end{figure}
\begin{figure}[t]
  \centering
  \includegraphics[width=0.8\linewidth]{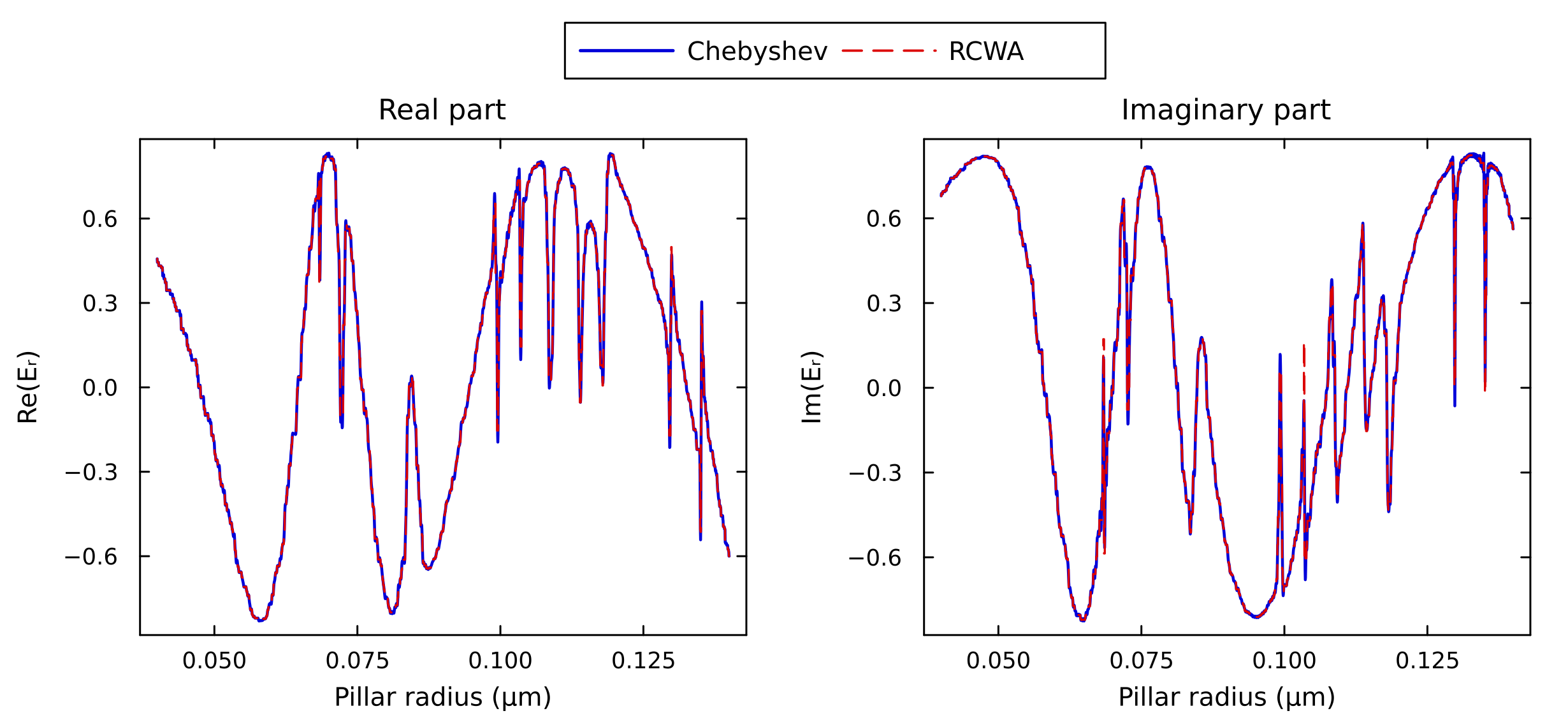}
  \caption{ Chebyshev-polynomial surrogate reconstructions (blue) of the real (left) and imaginary (right) parts of the transmitted electric field at normal incidence, compared with the Locally-Periodic Approximation (LPA) computed values (red) on which they were trained.}
  \label{fig:Fig2}
\end{figure}

The aim of this paper is to present a fast, differentiable far-field transform and to demonstrate its computational efficacy, rather than to advance the state of the art in metasurface transmission modeling, which we defer to future work. Accordingly, in our metalens design examples (Sec.~\ref{sec:normal}) we adopt the angle-independent model: we consider axisymmetric metalenses composed of circular dielectric pillars (Fig.~\ref{fig:Fig1}) whose transmission depends only on the pillar geometry and wavelength, $t(\mathbf{g}(r);\lambda)$. Within this model, we study single and cascaded configurations, with each pillar parametrized by a single radius $\rho(r)$; the two configurations offer different numbers of degrees of freedom for dispersion engineering. The surrogate functions for all normal-incidence configurations are fitted to pre-computed full-wave simulation data from sub-wavelength unit-cells (Fig.~\ref{fig:Fig2}). For our oblique-incidence studies (Sec.~\ref{sec:oblique}), where reliable angle-dependent models are not yet available, we instead optimize idealized, locally-applied phase mask profiles $t(r) = e^{i\varphi(r)}$ that isolates the performance of the far-field pipeline from the limitations of any particular meta-atom parametrization. These profiles represent an upper bound on what any lossless, axisymmetric metasurface, with angle insensitive or weakly-sensitive meta-atoms~\cite{wei2024spatially,wirth2025wide}, could achieve.

\section{Results and discussion}
\label{sec:3}
\subsection{Theoretical validations}
We validate our cylindrical far-field transform by computing the PSFs of a canonical near-field profile for which the expected behavior is known analytically: that of an \textit{ideal} lens whose phase exactly compensates the optical path from every aperture point $(r,\theta)$ to the focal spot at $(f\tan\alpha,\, 0,\, f)$:
\begin{align}
    \mathbf{E}_\text{ideal}(r,\theta) =
\exp{\left(
-i k
\left(
\sqrt{(r \cos{\theta} - f \tan{\alpha})^2 + r^2 \sin^2{\theta} + f^2}-f
\right)
\right)} \mathbf{\hat{y}}.
\end{align}

\begin{figure}[t]
  \centering
  \includegraphics[width=0.8\linewidth]{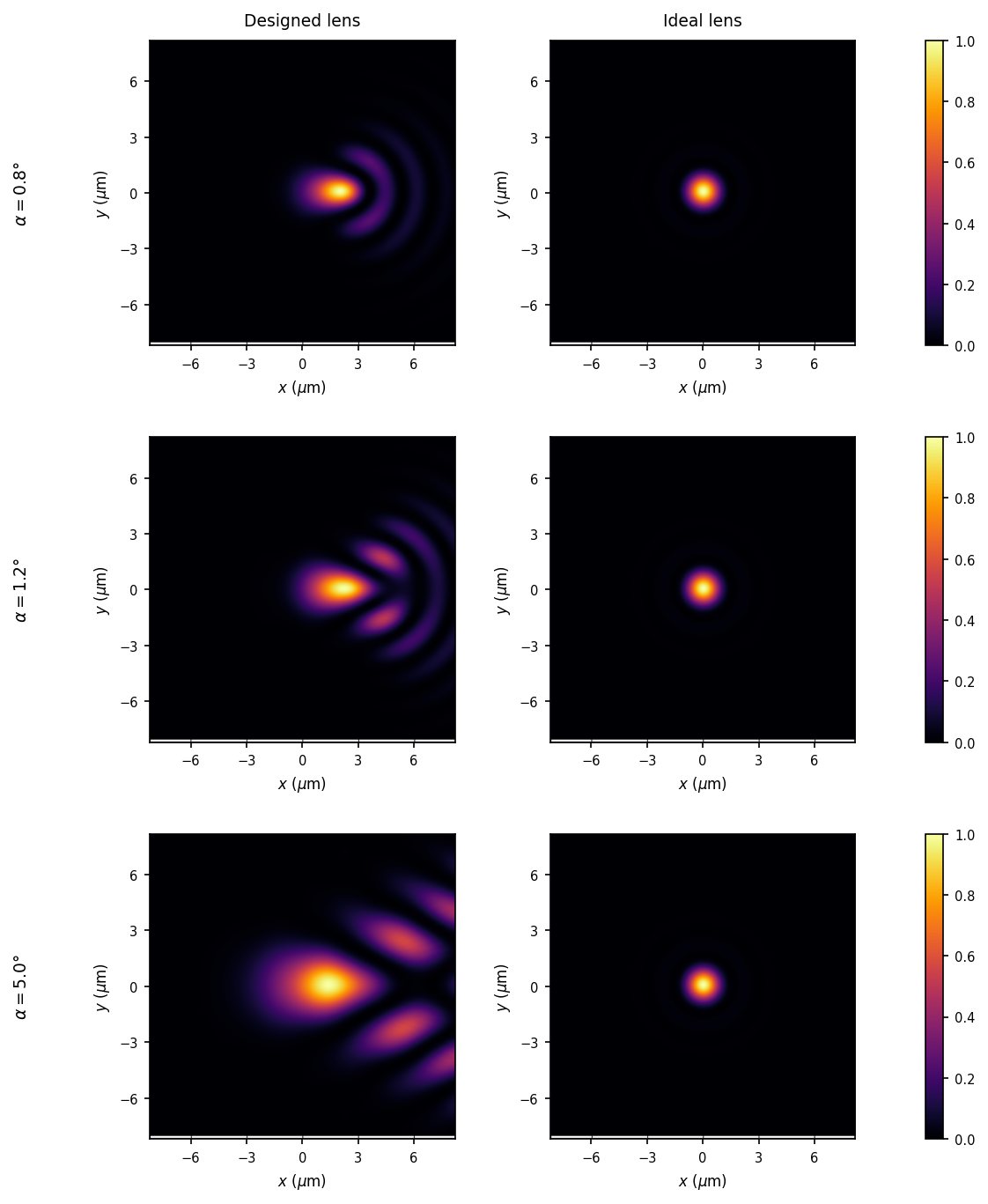}
  \caption{Heatmaps show far-field response of 4 mm designed lens for normal incident light (left column) and ideal lens (right column) at 500 nm to oblique angles of incidence at the shifted center ($f\times \mathrm{tan}\,\alpha$). From top to bottom the angles are 0.8, 1.2, and 5 degrees. }
  \label{fig:Fig3}
\end{figure}

This profile is $\theta$-dependent for $\alpha \neq 0$ and produces a diffraction-limited focal spot centered at $(f\tan\alpha,\, 0,\, f)$ for any incident angle. We project onto a single dominant polarization component $\hat{y}$ for simplicity. Fig.~\ref{fig:Fig3} compares the PSFs of the ideal and designed lens profiles across a range of incident angles for a lens with aperture diameter $8000\lambda$ and $\sim20{,}000$ radial sampling points. In the paraxial regime ($\alpha \lesssim 1^\circ$), both profiles yield nearly identical, diffraction-limited Airy disks centered at $x = f\tan\alpha$, consistent with paraxial shift invariance~\cite{Goodman2017}. As $\alpha$ increases, the designed-lens PSF degrades progressively due to uncorrected aberrations, while the ideal-lens PSF remains diffraction-limited, as expected. Despite computing the full focal-plane PSF, the normal-incidence computation took only 5.53 seconds, and the most oblique case ($\alpha = 5^\circ$, 885 $m$-modes processed in parallel across 32 threads) took 6.11 seconds. Both computations consumed well under 1~GB of RAM---orders of magnitude below the cost of alternative methods (Table~\ref{tab:timings}).

\begin{figure}[t]
  \centering
  \includegraphics[width=\textwidth]{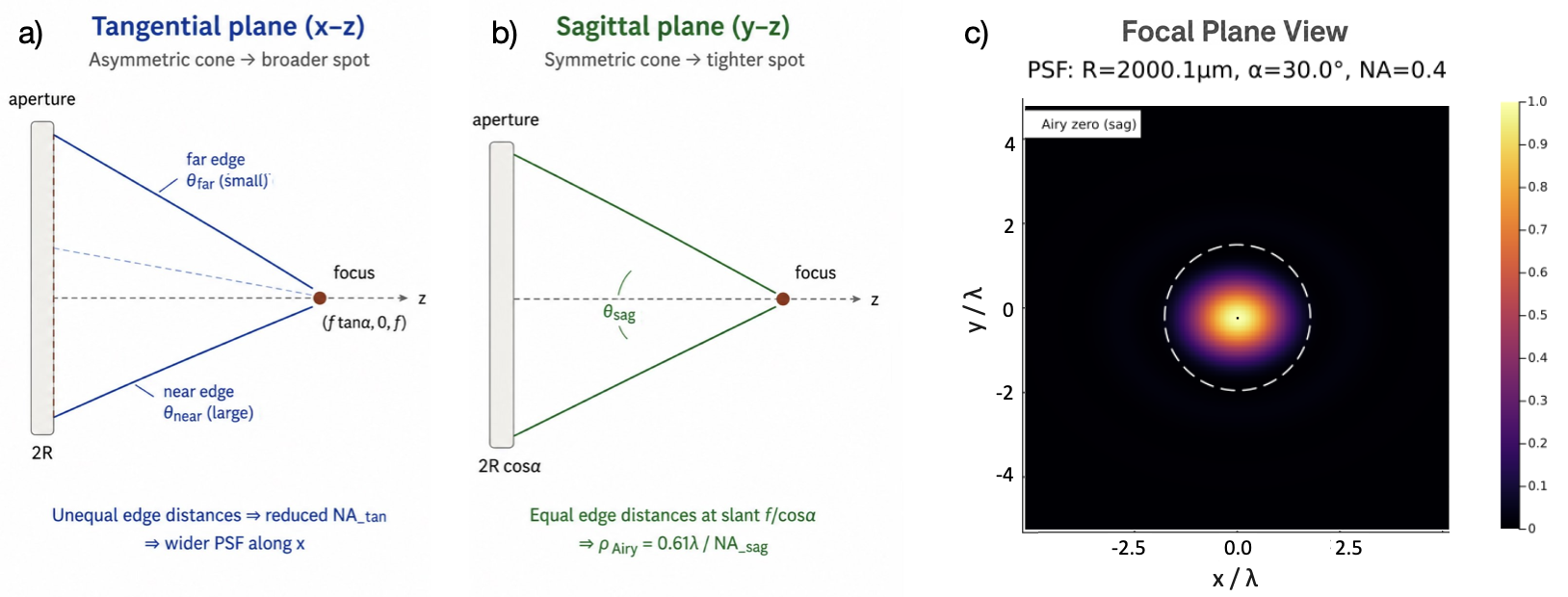}
  \caption{The figure shows PSF of the ideal oblique lens $\mathbf{E}_\text{ideal}(r,\theta)$ at $\alpha = 30^\circ$ ($R = 2$~mm, NA${} = 0.4$, $\lambda = 0.5$~$\mu$m). The $\theta$-dependent phase exactly compensates the optical path to the shifted focal spot at $(f\tan\alpha, 0, f)$. The focal spot is oval due to the different effective numerical apertures in the a) tangential ($x$) and b) sagittal ($y$) directions (illustrated in the schematics). c) The dashed circle marks the sagittal Airy zero $\rho_\text{Airy} = 0.61\lambda / \text{NA}_\text{sag}$; the sagittal cut matches this prediction to within 2\%, while the tangential extent is broader, consistent with the reduced tangential NA.}
  \label{fig:Fig4_new}
\end{figure}

At still larger angles, the $\theta$-dependence of $\mathbf{E}_\text{ideal}$ becomes essential and the PSF departs qualitatively from a circular Airy disk. Fig.~\ref{fig:Fig4_new} shows the ideal-lens PSF at $\alpha = 30^\circ$ for 4mm diameter, NA${} = 0.4$, and $\lambda = 0.5$~$\mu$m, which took only $\sim$4 secs on a 350-core dual-AMD EPYC machine. The focal spot is oval: narrower in the sagittal ($y$) direction and wider in the tangential ($x$) direction. This ellipticity arises because the shifted focal point at $x_0 = f\tan\alpha$ subtends different angular extents of the circular aperture in the two transverse planes. In the sagittal plane, the aperture spans its full radius $R$ at a slant distance $f/\cos\alpha$, giving an effective numerical aperture $\text{NA}_\text{sag} \approx R\cos\alpha / \sqrt{R^2\cos^2\alpha + f^2}$. In the tangential plane, the near and far edges of the aperture lie at unequal distances from the focal point, producing an asymmetric cone with a further-reduced effective NA. The dashed circle in Fig.~\ref{fig:Fig4_new} marks the sagittal Airy zero $\rho_\text{Airy} = 0.61\lambda / \text{NA}_\text{sag}$; the computed PSF matches this prediction to within 2\% along the sagittal cut, while the tangential extent exceeds it, consistent with the lower tangential NA. Despite requiring $M_\text{max} = 12{,}587$ azimuthal modes, the PSF is free of spurious aberrations, confirming the accuracy of the far-field transform at large oblique angles.

\subsection{RGB metalens designs at normal incidence}
\label{sec:normal}
We now combine our cylindrical far-field transform with metasurface transmission surrogates to optimize large-area polychromatic metalenses under normal incidence ($\alpha = 0$). Simple pillar meta-atoms offer substantially fewer degrees of freedom than freeform topology optimization~\cite{sun2025scalable}, making the optimization landscape prone to local minima. Consequently, both the choice of the objective function and the metalens initialization are essential for reliable optimization.

\begin{figure}[t]
  \centering
  \includegraphics[width=\textwidth]{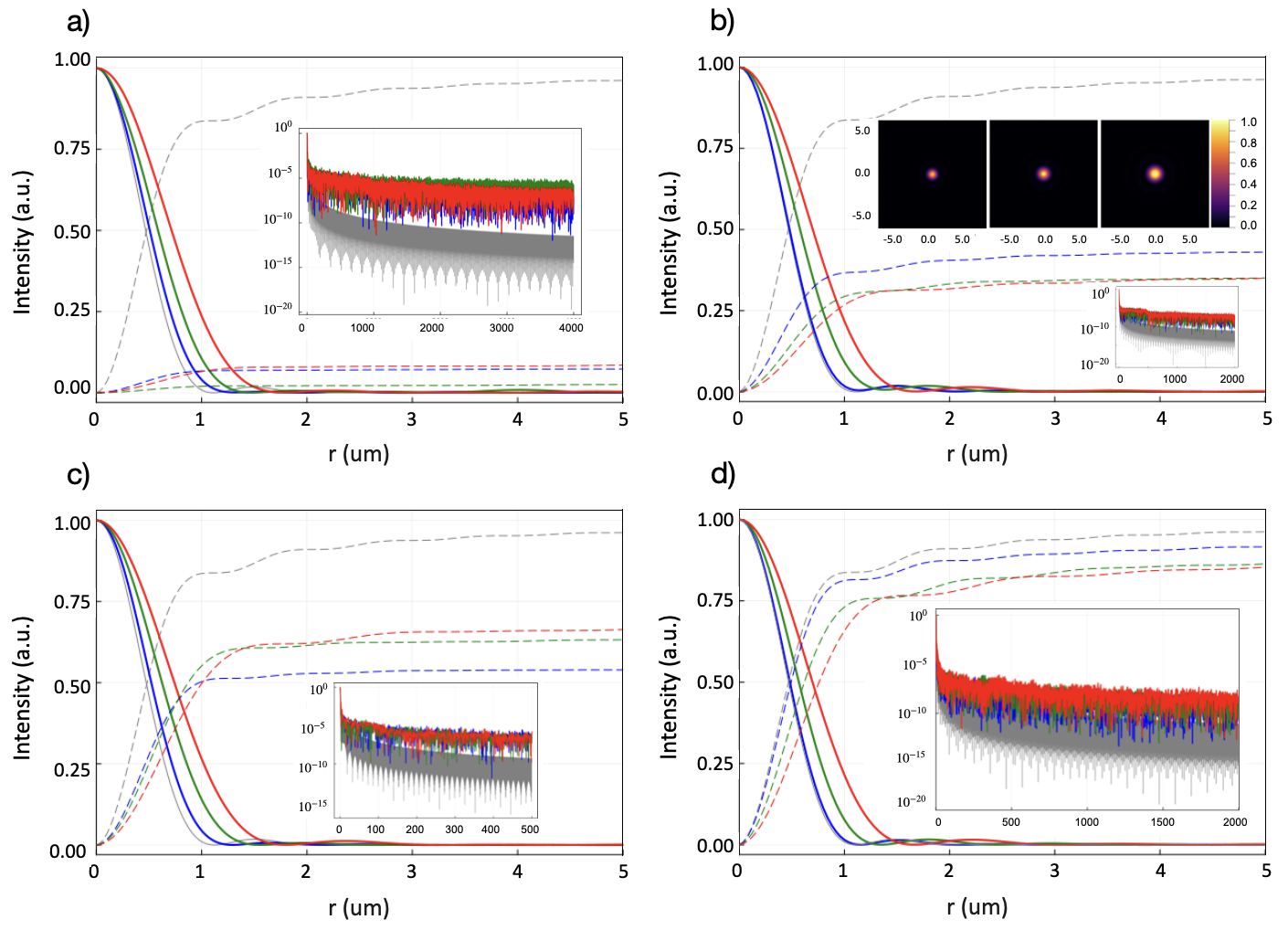}
  \caption{PSF radial cuts for multiple designed metalens structures with same specs, in linear (main plots) and log scales (insets). The blue, green and red curves show RGB results, and gray curves are ideal-lens profiles for blue, propagated to farfield plane (Solid lines for normalized intensities and dashed lines for enclosed power). Average absolute focusing efficiencies are $6\%$, $37.11\%$ , $50.93\%$ , $62.66\%$, respectively. a-c) Single metasurface, mono-pillar 4mm Si3N4 metalens designed using a) manually dispersion engineered near field and then ray traced to far field, b) direct optimization of far field using LPA and c) ZDA for same structure. d) Double-metasurface mono-pillar 4mm metalens with TiO2 pillars on the first surface and Si3N4 pillars on the second one, designed by direct optimization of far-field profile using LPA.}
  \label{fig:Fig5_new}
\end{figure}

\paragraph{Mono-pillar design.}
Fig.~\ref{fig:Fig5_new} shows the optimized PSFs of a 4~mm-diameter mono-pillar metalens (Si3N4-on-silica) at wavelengths 650~nm (red), 530~nm (green), and 446~nm (blue). Diffraction-limited focal spots are achieved at all three wavelengths using the multi-criteria objective
\begin{align}
    \mathcal{L} = \overline{\log \eta_\lambda} - A\;\sigma(\log \eta_\lambda) - B\;\Delta x_c,
\end{align}

which rewards the mean log focusing efficiency $\eta_\lambda$ across wavelengths while penalizing its standard deviation $\sigma$ (to promote achromatic balance) and the focal-spot centroid shift $\Delta x_c$ with $A=0.2$ and $B=100$. A gradient-free initialization of the pillar geometry proved essential for avoiding poor local minima.

\paragraph{Full-area PSF optimization.}
A distinguishing capability of our framework is the evaluation of the \emph{entire} PSF across the full focal plane---an analysis rarely reported in the metalens literature. Most studies present only a cropped central region, which may display a sharp Airy-like peak while concealing an elevated noise floor at larger radii. For example, a conventionally designed metalens based on mono-pillar meta-atoms---whose near fields were dispersion engineered and then ray-traced to the focal plane using Zemax---exhibits what appears to be a diffraction-limited focal spot; however, computing the full-area PSF reveals absolute focusing efficiencies as low as $\sim6\%$ (Fig.~\ref{fig:Fig5_new}a).  By comparison, the mono-pillar RGB metalens design, whose far-field PSFs are directly optimized using our differentiable cylindrical far-field transform and evaluated under the same efficiency metrics, achieves $\eta \approx 37.11\%$ (Fig.~\ref{fig:Fig5_new}b), which, though a significant improvement, still falls short of the ideal Airy disk efficiency ($>80\%$).  We attribute these limitations primarily to the simplicity of the meta-atom geometry and the locally periodic approximation, which together lack sufficient degrees of freedom to suppress the far-field noise floor. Overcoming these limitations likely requires more complex meta-atom geometries and transmission models beyond local sub-wavelength unit-cell approximations, e.g., supra-wavelength domain decomposition methods~\cite{lin2019overlapping} or the zoned discrete axisymmetry (ZDA) framework ~\cite{sun2025scalable}. Optimized lens using ZDA achieved 50.93\% (Fig.~\ref{fig:Fig5_new}c). Using our far-field transform in conjunction with full-wave ZDA, we have also optimized a 5cm-aperture mid-wave infrared (MWIR) metalens achieving 87\% absolute focusing efficiency; full details will be reported in a forthcoming publication.

\begin{figure}[H]
  \centering
  \includegraphics[width=\textwidth]{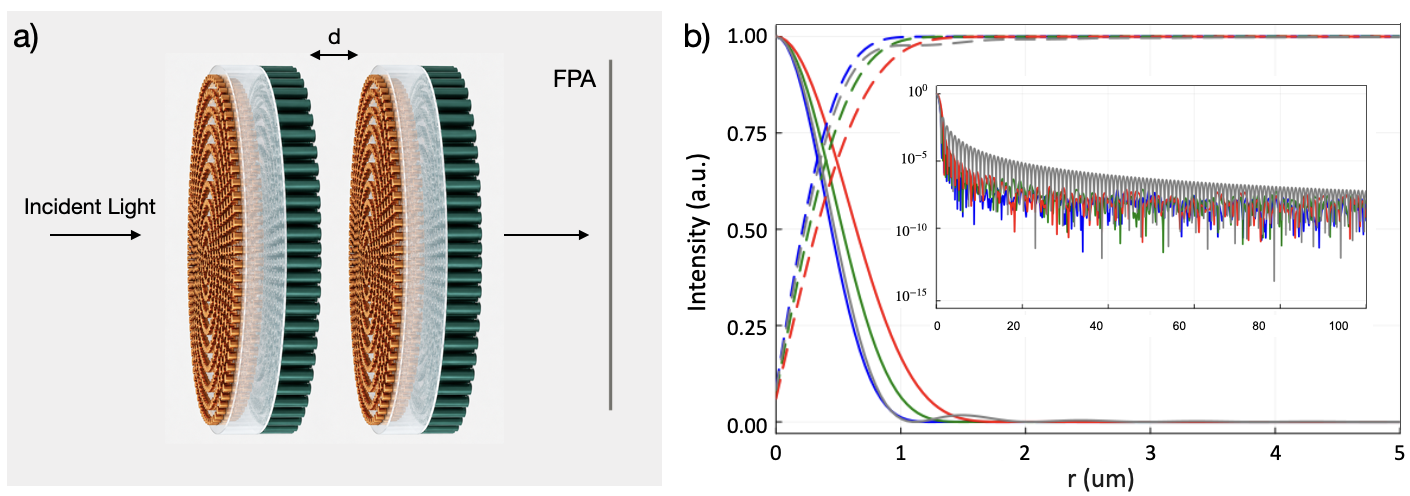}
  \caption{a) PSF radial cut for 4mm 4-metasurface structured cascaded metalens optimized for RGB  in linear (main plots) and log scales (insets), giving $96\%$ average relative focusing efficiency. The blue, green and red curves show RGB results, and gray curves are ideal-lens profiles for blue, propagated to farfield plane (Solid lines for normalized intensities and dashed lines for enclosed power). As shown in the schematic (b), light propagates through MS1, SiO2, MS2, air, MS3, SiO2, MS4, and air to reach the focal plane. MS1 \& MS3: 0.7 um long TiO2 Bi-pillars, bottom pillar embedded in SiO2. MS2 \& MS4: 0.8 um long Si3N4 mono-pillars.}
  \label{fig:Fig45_new}
\end{figure}

\paragraph{Cascaded design.}
To further enhance the absolute focusing efficiency, we combine two mono-pillar meta-atom designs on opposite sides of a single silica substrate (Fig.~\ref{fig:Fig1}), achieving an average absolute focusing efficiency of $62.66\%$ (Fig.~\ref{fig:Fig5_new}d).  Each three-wavelength PSF evaluation of this cascaded assembly takes approximately 0.082~seconds for 47{,}630 radial grid points (Table~\ref{tab:timings}), making full-stack gradient-based optimization computationally feasible. In order to further demonstrate the speed and potentials of our framework, a 4 metasurface-structure is optimized for RGB at normal-incidence giving average relative focusing efficiency of $96\%$ (Fig.~\ref{fig:Fig45_new}).  To the best of our knowledge, this constitutes the first demonstration of \textit{large-area} (several millimeters in diameter) cascaded metalens optimization without the use of ray tracing.

\subsection{Oblique incidence analysis}
\label{sec:oblique}
We now turn to oblique incidence, where the computational demands of the cylindrical far-field transform are greatest. Yet it is also where the structural advantage over Cartesian methods is most pronounced: the cylindrical cost is linear in the number of (embarrassingly parallel) angular momentum channels $M_{\text{max}}$, with $M_{\text{max}}\ll N_r$, whereas the 2D Cartesian FFT scales superlinearly in the radial grid size $N_r$ and limits parallel scaling. As noted in Sec.~\ref{sec:normal}, current locally periodic meta-atom models remain unreliable at large oblique angles: accurately training surrogate transmission functions $t(\mathbf{g}, \alpha, \beta, \lambda)$ over extended angular ranges is an open challenge (see Appendix~\ref{apx:LPA}), particularly for the tall, high-aspect-ratio pillars required for full $2\pi$ phase coverage. Rather than coupling our far-field transform to a potentially inaccurate meta-atom model, we instead validate the speed and scalability of our pipeline by optimizing \textit{proxy} radial phase-only transmission profiles $t(r) = e^{i\varphi(r)}$ that are independent of the meta-atom geometry. These idealized profiles represent the best-case scenario for a lossless, angle-independent metasurface, and they isolate the performance of the far-field transform from the limitations of any particular meta-atom model.

\begin{figure}[t]
  \centering
  \includegraphics[width=\linewidth]{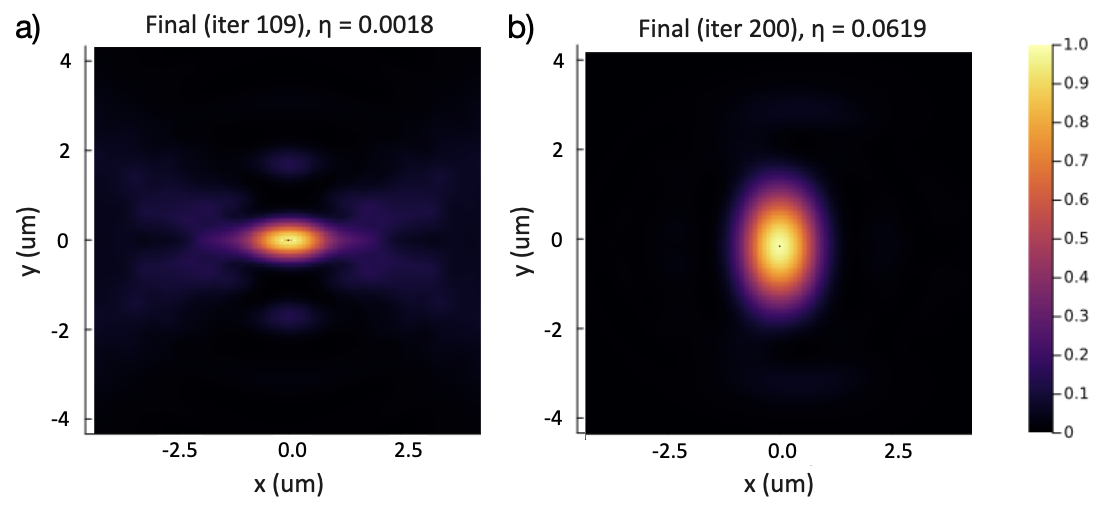}
  \caption{Optimized PSFs at $\alpha = 30^\circ$ oblique incidence ($\lambda = 0.5$~$\mu$m, NA${} = 0.4$). (a)~Single-surface design ($R = 2$~mm): the PSF is elongated along the tangential (horizontal) direction due to uncorrected coma. (b)~Doublet design ($R = 1$~mm, $d = 500$~$\mu$m): the inter-surface propagation provides $m$-dependent phase correction, partially correcting coma and yielding a $35\times$ higher focusing efficiency. The residual PSF elongation rotates to the sagittal (vertical) direction.}
  \label{fig:Fig7_new}
\end{figure}

\paragraph{Single-surface optimization.}
Our fast far-field transform evaluates the PSF of a 4~mm-diameter aperture ($R = 2$~mm, NA${} = 0.4$, $\lambda = 0.5$~$\mu$m) at $\alpha = 30^\circ$ in approximately 4~seconds, using 350 CPU threads to process $M_\text{max} = 12{,}587$ angular momentum modes in parallel over $N_r = 131{,}072$ radial grid points. This speed makes iterative optimization under oblique illumination practical. Fig.~\ref{fig:Fig7_new}(a) shows the optimized PSF for a $30^\circ$ oblique incidence angle. The optimization used adjoint-computed gradients to maximize the encircled power within the first null ring of a re-centered ideal Airy disk and converged in 109 iterations ($\sim$12~seconds per forward--adjoint iteration, $\sim$22~minutes total). The imperfect focal spot with residual aberrations are not a limitation of the optimizer or the far-field transform, but rather reflects a fundamental physical constraint: a rotationally symmetric surface applies the same radial phase $\varphi(r)$ to all angular momentum channels $m$, and therefore cannot correct coma, which dominates the aberration budget at large oblique angles.

\paragraph{Doublet optimization.}
A cascaded architecture with two surfaces separated by a distance $d$ overcomes this limitation. The inter-surface free-space propagation applies an $m$-dependent radial filter to each angular momentum channel---physically analogous to the aberration correction mechanism in thick refractive lens systems---enabling partial coma correction that is fundamentally inaccessible to a single surface. We optimize a doublet with a 2~mm diameter for both entrance and exit surfaces, separated by a 0.5~mm thick substrate, parameterized by two independent entrant and exit phase profiles $\varphi_1(r)$ and $\varphi_2(r)$. We note that our configuration is different from typical doublets where the entrance diameter is significantly smaller than the exit diameter, which helps separates the different incident angles. The doublet forward evaluation requires only $\sim$1.6~seconds (350 threads, $M_\text{max} = 6{,}303$ modes), and each forward--adjoint iteration takes $\sim$4.3~seconds. After 200 optimization iterations ($\sim$14~minutes), the doublet achieves a substantially improved focal spot with 35$\times$ more power concentrated in the Airy disk than in the case of the single-surface design (Fig.~\ref{fig:Fig7_new}(b)). Notably, the optimized doublet PSF is elongated along the sagittal (vertical) direction, in contrast to the singlet whose PSF is elongated tangentially (horizontally). This shape reversal is an artifact of our choice of optimization objective, which rewards the fraction of power that falls within a circular area around the central peak, without extra constraints on the PSF shape. The optimization effectively found that it could improve the encircled power fraction by correcting the dominant tangential aberration (partial coma correction), leaving the sagittal component as the leading residual.

These proxy-profile results demonstrate that our cylindrical far-field pipeline is fast enough to support full oblique-angle optimization of millimeter-scale metalens systems, including cascaded architectures, within minutes on a multi-core CPU. The integration of reliable angle-dependent metasurface transmission models, such as overlapping domain decomposition and/or axisymmetric maxwell solvers~\cite{sun2025scalable,Christiansen2021inverse,lin2021computational}, into this pipeline is straightforward, requiring only the replacement of $t(r) = e^{i\varphi(r)}$ with the accurate transmitted field for each azimuthal mode $\mathbf{E}_m(r)$.

\paragraph{Computational performance summary.}
Table~\ref{tab:timings} consolidates the computational cost of every task demonstrated in this paper. All timings were measured on a dual-socket AMD EPYC machine with 350 threads. For comparison, a single forward evaluation via oscillatory Green's function integrals requires on the order of 45~hours for a $\sim 10{,}000\lambda$ aperture~\cite{iserles2004quadrature}, and a two-dimensional FFT approach at the same resolution would exceed 1~TB of memory. Our cylindrical pipeline reduces these costs by three to four orders of magnitude, making not only single PSF evaluations but also iterative gradient-based optimization feasible within minutes.

\begin{table}[t]
\centering
\caption{Computational performance of the cylindrical far-field pipeline across all demonstrations in this paper. ``Fwd'' denotes a single forward PSF evaluation; ``Fwd+Adj'' includes the adjoint (gradient) pass. All timings on a 350-thread CPU.}
\label{tab:timings}
\begin{tabular}{@{}llrrrl@{}}
\toprule
Task & Aperture & $M_\text{max}$ & $N_r$ & Time & Note \\
\midrule
\multicolumn{6}{@{}l}{\textit{Sec.~\ref{sec:normal}: Normal incidence ($\alpha = 0$)}} \\
\quad Mono-pillar, 3$\lambda$ fwd & 4~mm & 1 & 47{,}630 & 0.027\,s & 3 wavelengths \\
\quad 2 Metasurfaces, 3$\lambda$ fwd & 4~mm & 1 & 47{,}630 & 0.082\,s & 3 wavelengths \\
\quad 4 Metasurfaces, 3$\lambda$ fwd & 4~mm & 1 & 47{,}630 & 1.043\,s & 3 wavelengths \\
\midrule
\multicolumn{6}{@{}l}{\textit{Sec.~\ref{sec:oblique}: Oblique incidence ($\alpha = 30^\circ$, $\lambda = 0.5$~$\mu$m)}} \\
\quad Singlet fwd & 4~mm & 12{,}587 & 131{,}072 & ${\sim}$4\,s & \\
\quad Singlet fwd+adj & 4~mm & 12{,}587 & 131{,}072 & ${\sim}$12\,s & \\
\quad Singlet optimization & 4~mm & 12{,}587 & 131{,}072 & ${\sim}$22\,min & 109 iters \\
\quad Doublet fwd & 2~mm & 6{,}303 & 131{,}072 & ${\sim}$1.6\,s & \\
\quad Doublet fwd+adj & 2~mm & 6{,}303 & 131{,}072 & ${\sim}$4.3\,s & \\
\quad Doublet optimization & 2~mm & 6{,}303 & 131{,}072 & ${\sim}$14\,min & 200 iters \\
\midrule
\multicolumn{6}{@{}l}{\textit{Reference methods (estimated)}} \\
\quad Green's function integral & $10{,}000\lambda$ & --- & --- & ${\sim}$45\,hr & \\
\quad 2D Cartesian FFT & 4~mm & --- & --- & $>$1\,TB RAM & infeasible \\
\bottomrule
\end{tabular}
\end{table}

\section{Conclusion and Outlook}
\label{sec:conclusion}
We have presented a fully differentiable cylindrical far-field transform that reduces the simulation and optimization of millimeter-scale axisymmetric metalenses from hours to seconds per iteration. By decomposing vectorial fields into angular momentum channels and applying an FFTLog-accelerated Hankel transform, the method replaces the two-dimensional Cartesian integral with embarrassingly parallel one-dimensional transforms per channel. Graf's addition theorem re-centers the focal plane at oblique incidence without synthesizing the full field, and a hand-derived adjoint of every pipeline stage enables gradient-based optimization at only ${\sim}65\%$ additional cost. We validated the pipeline against analytical Airy patterns, including the characteristic oval PSF of ideal lens at $30^\circ$ and demonstrated three applications. The first was a polychromatic metalens design at normal incidence, where full-area PSF evaluation revealed that existing designs achieve far lower efficiencies than commonly reported. The second, a single-surface oblique optimization, which confirmed the fundamental coma limit of rotationally symmetric phase profiles. And last, a cascaded doublet optimization at oblique incidence, which achieved $35\times$ higher focusing efficiency through $m$-dependent inter-surface phase correction. To our knowledge, the latter is the first such result without ray tracing at the scale of several millimeters in diameter aperture.

Several directions for future work follow naturally. First, integrating angle-resolved metasurface transmission models $t(\mathbf{g}, \alpha, \beta; \lambda)$ into the pipeline is architecturally straightforward: the local incidence angles at each meta-atom can be extracted analytically from the VCH expansion (Appendix~\ref{apx:LPA}), and the only required change is replacing the proxy phase profile $e^{i\varphi(r)}$ with the appropriate surrogate. The bottleneck is both physical and computational: tall, high-aspect-ratio pillars support sharp angular resonances, making accurate surrogate training over extended angular ranges an open problem. Recent work~\cite{zerbe2026analytical} points to promising directions. Second, the framework generalizes immediately to multi-surface stacks with arbitrary inter-surface separations and different aperture radii per surface; each additional surface adds one Hankel round-trip per mode at the same per-mode cost. Third, full-area PSFs computed by our pipeline are directly compatible with end-to-end co-optimization of metalens optics and computational back-ends (e.g., Wiener deconvolution or learned reconstruction networks), enabling joint design of the physical optic and the digital post-processing. While our current implementation targets shared-memory multi-core CPUs, the per-channel independence of the pipeline maps naturally onto GPU architectures, where the thousands of modes at large oblique angles would saturate modern hardware and yield additional speedups. By making full-area PSF evaluation fast and differentiable at millimeter-to-centimeter scale, our approach removes a longstanding bottleneck in metalens inverse design--enabling optimization in regimes where wave-optics methods were previously intractable and ray tracing too approximate.

\appendix
\section{Vector Cylindrical Harmonics}
\label{apx:vch}

We derive the vector cylindrical harmonic (VCH) basis functions used in the main text. The approach follows Stratton~\cite{Stratton1941}: introduce $z$-directed scalar potentials that satisfy the Helmholtz equation, then obtain the transverse field components by differentiation.

\paragraph{Scalar Helmholtz equation.}
In a source-free, homogeneous medium with wavenumber $k = \omega\sqrt{\mu\varepsilon}$, both the electric and magnetic Hertz potentials reduce to solving the scalar Helmholtz equation:
\begin{equation}
\nabla^2\psi + k^2\psi = 0.
\label{eq:helmholtz}
\end{equation}
In cylindrical coordinates $(r,\theta,z)$, this becomes
\begin{equation}
\frac{1}{r}\frac{\partial}{\partial r}\!\left(r\frac{\partial\psi}{\partial r}\right) + \frac{1}{r^2}\frac{\partial^2\psi}{\partial\theta^2} + \frac{\partial^2\psi}{\partial z^2} + k^2\psi = 0.
\end{equation}
Separation of variables $\psi = R(r)\,\Theta(\theta)\,Z(z)$ yields three eigenvalue equations. The azimuthal equation gives $\Theta(\theta) = e^{im\theta}$ with integer $m$; the axial equation gives $Z(z) = e^{ik_z z}$ with $k_z^2 = k^2 - k_r^2$; and the radial equation is Bessel's equation of order $m$:
\begin{equation}
r\frac{d}{dr}\!\left(r\frac{dR}{dr}\right) + (k_r^2 r^2 - m^2)\,R = 0,
\end{equation}
with regular solution $R(r) = J_m(k_r r)$. The elementary scalar eigenfunctions are therefore
\begin{equation}
\psi_{m,k_r}(r,\theta,z) = J_m(k_r r)\,e^{im\theta}\,e^{ik_z z}.
\label{eq:psi_mk}
\end{equation}

\paragraph{TM modes.}
Introducing the magnetic vector potential $\mathbf{A} = \hat{z}\,\mu\psi$ and computing $\mathbf{H} = \frac{1}{\mu}\nabla\times\mathbf{A}$, $\mathbf{E} = \frac{1}{-i\omega\varepsilon}\nabla\times\mathbf{H}$, we obtain the TM (transverse magnetic, $H_z = 0$) field components. The transverse electric field of a TM mode with indices $(m, k_r)$ is
\begin{align}
\mathbf{E}_{m,k_r}^{\text{TM}}(r) &= \frac{k_z}{\omega\varepsilon}\left( k_r J_m'(k_r r)\,\hat{r} + \frac{im}{r}J_m(k_r r)\,\hat{\theta} \right), \label{eq:ETM}
\end{align}
where $J_m'(x) \equiv dJ_m/dx$.

\paragraph{TE modes.}
Introducing the electric vector potential $\mathbf{F} = \hat{z}\,\varepsilon\psi$ and computing $\mathbf{E} = -\frac{1}{\varepsilon}\nabla\times\mathbf{F}$, we obtain the TE (transverse electric, $E_z = 0$) field components:
\begin{align}
\mathbf{E}_{m,k_r}^{\text{TE}}(r) &= -\frac{im}{r}J_m(k_r r)\,\hat{r} + k_r J_m'(k_r r)\,\hat{\theta}. \label{eq:ETE}
\end{align}

\paragraph{Summary.}
Equations~\eqref{eq:ETM} and~\eqref{eq:ETE} are the transverse VCH basis functions $\mathbf{E}_{m,k_r}^{\text{TE,TM}}(r)$ appearing in the main text (Eq.~\ref{eq:Enear}). Each basis function is characterized by azimuthal order $m$ and radial wavenumber $k_r$, with axial propagation factor $e^{ik_z z}$ where $k_z = \sqrt{k^2 - k_r^2}$. The key property exploited by our pipeline is that the Hankel-transform integrals (Eqs.~\ref{eq:ATE}--\ref{eq:ATM}) involve only one-dimensional integrals over $r$ for each $m$-channel.
\section{Mode expansion coefficients in VCH basis}
\label{apx:coefs}
We derive the projection formulas (Eqs.~\ref{eq:ATE}--\ref{eq:ATM}) that extract the TE and TM expansion coefficients from a given near field. The key ingredient is the orthogonality of the VCH basis functions under the inner product weighted by $r$.

\paragraph{TE--TE inner product.}
Taking the dot product of $\mathbf{E}_{m,k_r}^{\text{TE}}(r)$ (Eq.~\ref{eq:ETE}) with $\mathbf{E}_{m,k_r'}^{\text{TE}*}(r)$ gives
\begin{equation}
\mathbf{E}_{m,k_r}^{\text{TE}} \cdot \mathbf{E}_{m,k_r'}^{\text{TE}*} = \frac{m^2}{r^2}\,J_m(k_r r)\,J_m(k_r' r) + k_r k_r'\,J_m'(k_r r)\,J_m'(k_r' r).
\label{eq:TETE_dot}
\end{equation}
Using the Bessel recurrence relations
\begin{equation}
\frac{2m}{x}\,J_m(x) = J_{m-1}(x) + J_{m+1}(x), \qquad 2\,J_m'(x) = J_{m-1}(x) - J_{m+1}(x),
\end{equation}
to rewrite the $m/r$ and derivative terms, and expanding the products, the cross terms cancel and we obtain
\begin{equation}
\mathbf{E}_{m,k_r}^{\text{TE}} \cdot \mathbf{E}_{m,k_r'}^{\text{TE}*} = \frac{k_r k_r'}{2}\Big[J_{m+1}(k_r r)\,J_{m+1}(k_r' r) + J_{m-1}(k_r r)\,J_{m-1}(k_r' r)\Big].
\end{equation}
Integrating over $r$ with the Bessel orthogonality relation
\begin{equation}
\int_0^\infty r\,J_m(k_r r)\,J_m(k_r' r)\,dr = \frac{\delta(k_r - k_r')}{k_r},
\label{eq:bessel_ortho}
\end{equation}
yields
\begin{equation}
\int_0^\infty \mathbf{E}_{m,k_r}^{\text{TE}} \cdot \mathbf{E}_{m,k_r'}^{\text{TE}*}\;r\,dr = k_r\,\delta(k_r - k_r').
\end{equation}
Since the TE and TM basis functions are orthogonal ($\int_0^\infty \mathbf{E}_{m,k_r}^{\text{TE}} \cdot \mathbf{E}_{m,k_r'}^{\text{TM}*}\;r\,dr = 0$), projecting the near-field expansion (Eq.~\ref{eq:Enear}) onto $\mathbf{E}_{m,k_r}^{\text{TE}*}$ isolates the TE coefficient:
\begin{equation}
A_m^{\text{TE}}(k_r) = \frac{1}{k_r}\int_0^\infty \mathbf{E}_m^{\text{Near}}(r) \cdot \mathbf{E}_{m,k_r}^{\text{TE}*}(r)\;r\,dr,
\end{equation}
where the factor of $\frac{1}{k_r}$ arises from the normalization $k_r\,\delta(k_r - k_r')$ and the convention in Eq.~\ref{eq:Enear}. The azimuthal integral contributes $2\pi\,\delta_{m,m'}$ and has already been absorbed into the definition of $\mathbf{E}_m^{\text{Near}}(r)$.

\paragraph{TM coefficients.}
An analogous calculation using the TM basis functions (Eq.~\ref{eq:ETM}) yields
\begin{equation}
\int_0^\infty \mathbf{E}_{m,k_r}^{\text{TM}} \cdot \mathbf{E}_{m,k_r'}^{\text{TM}*}\;r\,dr = \frac{k_z^2}{\omega^2\varepsilon^2}\,k_r\,\delta(k_r - k_r'),
\end{equation}
giving the TM projection formula
\begin{equation}
A_m^{\text{TM}}(k_r) = \frac{\omega^2\varepsilon^2}{\,k_z^2k_r}\int_0^\infty \mathbf{E}_m^{\text{Near}}(r) \cdot \mathbf{E}_{m,k_r}^{\text{TM}*}(r)\;r\,dr.
\end{equation}

\section{Local-field expansion in azimuthal harmonics}
\label{apx:pwe}
For an axisymmetric metalens with transmission function $t(r)$ under an obliquely incident plane wave (angle $\alpha$ in the $xz$-plane, $\hat{y}$-polarized), the transmitted near field is
\begin{equation}
\mathbf{E}_\text{Near}(r,\theta) = t(r) \exp\!\left(i k_x\, r \cos\theta \right)\,\hat{y},
\label{eq:Enear_oblique}
\end{equation}
where $k_x = k\sin\alpha$ is the transverse component of the incident wavevector. The azimuthal decomposition of this field can be carried out analytically using the Jacobi-Anger expansion:
\begin{equation}
e^{i k_x r \cos\theta} = \sum_{m=-\infty}^{\infty} i^m\, J_m(k_x r)\, e^{im\theta}.
\label{eq:jacobi_anger}
\end{equation}
Substituting into Eq.~\eqref{eq:Enear_oblique}, the $m$-th azimuthal component of the near field is
\begin{equation}
\mathbf{E}_m^{\text{Near}}(r) = i^m\, t(r)\, J_m(k_x r)\,\hat{y}.
\label{eq:Em_near}
\end{equation}
This expression bypasses the need for a numerical FFT in $\theta$ and directly provides the radial input to the Hankel-transform pipeline (Eqs.~\ref{eq:ATE}--\ref{eq:ATM}) for each $m$-channel. In practice, the sum is truncated at $|m| \leq M_\text{max}$, where $M_\text{max} \approx k_x R + \mathcal{O}(k_x R)^{1/3}$ is set by the evanescent decay of $J_m(k_x r)$ for $|m| \gg k_x R$, with $R$ the aperture radius. For normal incidence ($\alpha = 0$), $k_x = 0$ and $J_m(0) = \delta_{m,0}$, so only $m = 0$ contributes---or, for a $\hat{y}$-polarized plane wave decomposed into TE and TM channels, only $m = \pm 1$.

\section{Efficient computation of negative-$m$ harmonics under oblique incidence}
\label{apx:negm}
For a $\hat{y}$-polarized plane wave incident in the $xz$-plane, the transmitted near field $\mathbf{E}_\text{Near}(r,\theta)$ possesses a mirror symmetry $\theta \to -\theta$ (equivalently, $y \to -y$). This symmetry relates the negative-$m$ azimuthal components to their positive-$m$ counterparts, halving the number of independent Hankel transforms.

Specifically, from Eq.~\eqref{eq:Em_near} and the Bessel identity $J_{-m}(x) = (-1)^m J_m(x)$, we have
\begin{equation}
\mathbf{E}_{-m}^{\text{Near}}(r) = (-1)^m\,\mathbf{E}_{m}^{\text{Near}}(r).
\end{equation}
Since the Hankel transform and propagation phase $e^{ik_z z}$ are both independent of the sign of $m$, the spectral coefficients inherit the same symmetry:
\begin{equation}
A_{-m}^{\text{TE,TM}}(k_r) = (-1)^m\, A_{m}^{\text{TE,TM}}(k_r).
\end{equation}
In practice, we compute the forward pipeline only for $m \geq 0$ and reconstruct the negative-$m$ coefficients using this relation, yielding an additional $2\times$ speedup beyond the parallelization over $m$-channels.

\section{Focal-plane re-centering via Graf's addition theorem}
\label{apx:graf}

For oblique incidence at angle $\alpha$, the focal spot is displaced to $(x_0, 0, f)$ with $x_0 = f\tan\alpha$. The propagated field in the global frame is
\begin{equation}
  u(r,\theta,f) = \sum_n \left[\int_0^k \tilde{A}_n(k_r)\,J_n(k_r r)\,k_r\,dk_r\right] e^{in\theta},
\end{equation}
where $\tilde{A}_n(k_r) = A_n(k_r)\,e^{ik_z f}$ are the propagated spectral coefficients. To evaluate $u$ in a local coordinate system $(\rho,\psi)$ centered at the shifted focus, where $x = x_0 + \rho\cos\psi$ and $y = \rho\sin\psi$, we apply Graf's addition theorem~\cite{Watson1944,lozier2003nist}. For $\rho < x_0$, each global basis function can be re-expanded as
\begin{equation}
  J_n(k_r r)\,e^{in\theta} = \sum_{m=-\infty}^{\infty} J_m(k_r x_0)\,J_{n-m}(k_r \rho)\,e^{i(n-m)\psi}.
  \label{eq:graf_theorem}
\end{equation}
Substituting into the far-field sum and relabeling $l = n - m$ (so $n = m + l$) yields
\begin{equation}
  u(\rho,\psi,f) = \sum_l \left[\int_0^k B_l(k_r)\,J_l(k_r \rho)\,k_r\,dk_r\right] e^{il\psi},
\end{equation}
where the local modal coefficients are given by the mode convolution
\begin{equation}
  B_l(k_r) = \sum_{|m| \leq M_\text{max}} \tilde{A}_{m+l}(k_r)\,J_m(k_r x_0), \qquad |l| \leq L_{\text{max}}.
  \label{eq:Bl_graf}
\end{equation}
The local coefficients $B_l$ feed directly into the inverse Hankel transform (Step~5), exactly as in the normal-incidence case.

Two properties make this step computationally efficient. First, $J_m(k_r x_0)$ decays exponentially for $|m| > k_r x_0$, so the effective summation length at each $k_r$ is $\min(M_\text{max},\,\lceil k_r x_0 \rceil + 20)$ rather than the full $2M_\text{max}+1$. Second, only a small number of local modes $L_{\text{max}} \approx k \cdot \text{NA} \cdot \rho_\text{max}$ are needed to resolve the PSF (e.g., $L_{\text{max}} \approx 15$ for $\rho_\text{max} = 5\lambda$), so the convolution produces far fewer outputs than inputs. The Bessel weights $J_0(k_r x_0), \ldots, J_{M_\text{max}}(k_r x_0)$ are computed via Miller's backward recurrence, which is approximately $10\times$ faster than evaluating each order independently, and the identity $J_{-m} = (-1)^m J_m$ halves the number of evaluations. The inner loop over $m$ is vectorized with SIMD instructions, and the outer loop over $k_r$ is parallelized across threads.

At normal incidence ($\alpha = 0$), $x_0 = 0$ and $J_m(0) = \delta_{m,0}$, so the Graf shift reduces to the identity $B_l = \tilde{A}_l$ and can be skipped entirely. Importantly, the local grid $\rho$ can extend to any desired radius---the Graf shift does not restrict the output region, it simply avoids the cost of synthesizing the full global focal plane when only a neighborhood of the focus is needed.

\section{Scalable backpropagation through cylindrical far-field transform}
\label{apx:backprop}
To enable gradient-based optimization of loss functionals $\mathcal{L}(\mathbf{E}_\text{Far})$, we require the adjoint (reverse-mode derivative) of the full cylindrical far-field pipeline. The pipeline consists of a sequence of linear and pointwise operations---azimuthal decomposition, Hankel transforms, propagation phase multiplication, Graf addition, inverse Hankel transforms, and angular synthesis---each of which admits a straightforward adjoint.

The key observation is that the Hankel transform is \textit{self-adjoint}: the adjoint of the forward FFTLog (which computes $A_m(k_r) = \mathcal{H}_m[r\,u_m](k_r)/k_r$) is an FFTLog of the same order $m$ applied to the incoming cotangent, but with the conjugated filter kernel $\overline{U_m(q)}$ in place of $U_m(q)$, where $U_m(q)$ is the Mellin-space representation of the Bessel function~\cite{Hamilton2000}. The diagonal weights ($r$, $1/k_r$, and their inverse-transform counterparts $k_r$, $1/\rho$) are real pointwise multiplications, so their adjoints are the same multiplications applied to the cotangent, as in Algorithm~\ref{alg:adj_single}. Since $|U_m(q)| = 1$ for all $q$ (a consequence of the Gamma function reflection formula), the adjoint FFTLog has identical numerical stability to the forward.

The adjoint of the propagation phase $e^{ik_z z}$ is multiplication by its conjugate $e^{-ik_z z}$. The adjoint of the Graf addition theorem (a mode convolution $B_l = \sum_m \tilde{A}_{m+l}\,J_m(k_r x_0)$) is a transposed convolution with the same Bessel weights. These adjoints are implemented as fused, multithreaded operations that mirror the forward pipeline's parallelism over $k_r$ and $m$-channels, achieving comparable throughput.

In our implementation, the forward and adjoint steps are fused in pairs (propagation + Graf shift, and FFTLog + mode construction) to avoid materializing large intermediate arrays, reducing peak memory by over 50~GB at production scale. The total adjoint cost is $\sim$65\% of the forward evaluation.

\section{Locally periodic metasurface transmission model}
\label{apx:LPA}

In this work, we model each metasurface layer using the locally periodic approximation (LPA), in which the response of each meta-atom is treated as if it were embedded in an infinite periodic array of identical elements. Under this approximation, the transmission at position $(r,\theta)$ depends only on the local meta-atom geometry $\mathbf{g}$ (as shown in Fig.~\ref{fig:Fig1}b-c) and, in the fully angle-resolved case, on the local angles of incidence $\alpha_{\text{loc}}$ and $\beta_{\text{loc}}$ as experienced by that meta-atom:
\begin{align}
    \mathbf{E}_\text{out}(r,\theta) &= t(r,\theta;\lambda)\, \mathbf{E}_\text{in}(r,\theta), \\
    t(r,\theta;\lambda) &= t\bigl(\mathbf{g}(r,\theta),\,\alpha_{\text{loc}}(r,\theta),\,\beta_{\text{loc}}(r,\theta);\,\lambda\bigr).
\end{align}
Our normal-incidence designs (Sec.~\ref{sec:normal}) use an angle-independent simplification $t(\mathbf{g};\lambda)$. Here we outline the full angle-resolved formulation, which becomes necessary for oblique illumination and cascaded architectures where the inter-surface field arrives at nontrivial local angles.

\paragraph{Inferring local angles from the incident field.}
The local angles of incidence are not externally specified. They must be inferred from the incident wave field itself. We adopt an eikonal (ray-optics) approximation: writing $\mathbf{E}_\text{in} = |\mathbf{E}_\text{in}|\,e^{i\varphi}$, the local propagation direction at each meta-atom is
\begin{equation}
    \hat{\mathbf{u}}(r,\theta) = \hat{\nabla}_{x',y'}\, \varphi,
\end{equation}
where $\hat{\nabla}$ denotes the unit gradient and $(x',y')$ is a local Cartesian frame aligned with the unit cell of the meta-atom at global position $(r,\theta)$. For an axisymmetric metalens, the natural cell orientation has $x'$ along the radial direction $\hat{r}$ and $y'$ along the azimuthal direction $\hat{\theta}$. The transformation from global Cartesian $(x,y) = (r\cos\theta,\, r\sin\theta)$ to the local cell frame is then a rotation by $-\theta$:
\begin{equation}
    \begin{pmatrix} x' \\ y' \end{pmatrix}
    = \begin{pmatrix} \cos\theta & \sin\theta \\ -\sin\theta & \cos\theta \end{pmatrix}
    \begin{pmatrix} x \\ y \end{pmatrix}.
    \label{eq:cell_rotation}
\end{equation}
Under this rotation, the local gradient operator reduces to the cylindrical gradient components:
\begin{equation}
    \nabla_{x',y'} = \left(\frac{\partial}{\partial r},\; \frac{1}{r}\frac{\partial}{\partial\theta}\right),
    \label{eq:local_grad}
\end{equation}
since $\partial_{x'} = \cos\theta\,\partial_x + \sin\theta\,\partial_y = \partial_r$ and $\partial_{y'} = -\sin\theta\,\partial_x + \cos\theta\,\partial_y = r^{-1}\partial_\theta$. This coordinate transformation is exact and avoids any numerical resampling.

The local incidence angles are then extracted from the unit gradient of the phase. The phase gradient can be evaluated without finite differences via the log-derivative identity
\begin{equation}
    \hat{\nabla}\,\varphi
    = i\,\hat{\nabla}\log|\mathbf{E}_\text{in}\cdot\hat{\mathbf{p}}|
    - i\,\hat{\nabla}\log(\mathbf{E}_\text{in}\cdot\hat{\mathbf{p}}),
\end{equation}
where $\hat{\mathbf{p}}$ is a reference polarization direction. Specifically, the local polar angle of incidence $\alpha_\text{loc}$ and azimuthal angle $\beta_\text{loc}$ at each meta-atom are
\begin{equation}
    \alpha_\text{loc} = \arcsin\!\left(\frac{|\nabla_{x',y'}\varphi|}{k}\right), \qquad
    \beta_\text{loc} = \operatorname{atan2}\!\left(\frac{1}{r}\frac{\partial\varphi}{\partial\theta},\; \frac{\partial\varphi}{\partial r}\right).
\end{equation}

As a consistency check, consider a plane wave incident at angle $\alpha_0$ in the $xz$-plane on the first surface, giving $\varphi = k\sin\alpha_0\, r\cos\theta$. The local gradient components are $\partial_r\varphi = k\sin\alpha_0\cos\theta$ and $r^{-1}\partial_\theta\varphi = -k\sin\alpha_0\sin\theta$, so $|\nabla\varphi| = k\sin\alpha_0$ everywhere and $\alpha_\text{loc} = \alpha_0$, recovering the global incidence angle as expected. The azimuthal angle $\beta_\text{loc} = -\theta$ rotates with position, reflecting the fact that each radially-oriented cell sees the tilt from a different direction.

\paragraph{Analytical gradient from the VCH expansion.}
In a cascaded metalens stack, the incident field on each downstream surface is the far field of the preceding surface, which is already available in the VCH basis. The spatial gradient (Eq.~\ref{eq:local_grad}) can therefore be computed analytically by differentiating the VCH expansion term by term:
\begin{equation}
    \nabla_{x',y'}\bigl(\mathbf{E}_\text{in}\cdot\hat{\mathbf{p}}\bigr)
    = \sum_{m}\sum_{n \in \{\text{TE,TM}\}} \int dk_r\; A_m^n(k_r)\,
    \nabla_{x',y'}\!\left[
    \bigl(\mathbf{E}_{m,k_r}^n(r)\cdot\hat{\mathbf{p}}\bigr)\,
    e^{im\theta}\,e^{ik_z z}
    \right].
\end{equation}
The derivatives of $J_m(k_r r)\,e^{im\theta}$ with respect to $r$ and $\theta$ are standard Bessel recurrences, so the local incidence angles at every meta-atom are obtained as a byproduct of the far-field evaluation at negligible additional cost. We leave the detailed implementation and validation of this coupled angle-resolved pipeline to future work.

\paragraph{Surrogate model challenges.}
Employing the full angle-resolved model requires training high-dimensional surrogate functions $t(\mathbf{g},\alpha,\beta,\lambda)$ from pre-computed full-wave simulations. While a range of surrogate approaches have been developed, from polynomial interpolation to neural networks~\cite{Ma2021deep,So2020deep,AnNeuralAdj2020}, accurately fitting surrogates for tall, high-aspect-ratio meta-atom geometries over extended angular and spectral ranges remains an open challenge. A potentially viable strategy is to partition each input dimension into $k$ intervals, yielding $k^d$ independent surrogates each trained over a small hypercube in parameter space. We defer a systematic investigation to future work.

\section*{Acknowledgement}
A.K., A.S., W.T., J.P., Q.W., W.-T.C. and Z.L. were supported by the US Department of the Navy (DON) under Contracts No. N6833525C0134, N6833526C0159, N6833523C0545 and N6833525C0011. A.K., A.S., W.T., and Z.L. were supported in part by the US Army Research Office (ARO) under Contract No. W911NF2410390. The authors extend their gratitude to the NAVAIR TPOCs for their invaluable support and guidance
throughout this project under the DON Contract N6833525C0134.

\bibliographystyle{unsrt}  
\bibliography{references}

@article{Groever2017,
  author  = {Groever, Benedikt and Chen, Wei Ting and Capasso, Federico},
  title   = {Meta-Lens Doublet in the Visible Region},
  journal = {Nano Letters},
  volume  = {17},
  number  = {8},
  pages   = {4902--4907},
  year    = {2017},
  doi     = {10.1021/acs.nanolett.7b01888}
}

@article{Pestourie2018,
  author  = {Pestourie, Rapha{\"e}l and P{\'e}rez-Arancibia, Carlos and Lin, Zin and Shah, Wonseok and Capasso, Federico and Johnson, Steven G.},
  title   = {Inverse Design of Large-Area Metasurfaces},
  journal = {Optics Express},
  volume  = {26},
  number  = {26},
  pages   = {33732--33747},
  year    = {2018},
  doi     = {10.1364/OE.26.033732}
}

@book{Goodman2017,
  author    = {Goodman, Joseph W.},
  title     = {Introduction to {F}ourier Optics},
  edition   = {4th},
  publisher = {W.~H.~Freeman},
  year      = {2017}
}

@book{Stratton1941,
  author    = {Stratton, Julius Adams},
  title     = {Electromagnetic Theory},
  publisher = {McGraw-Hill},
  year      = {1941}
}

@article{beckman2024nonuniform,
  title={A Nonuniform Fast Hankel Transform},
  author={Beckman, Paul G and O'Neil, Michael},
  journal={arXiv preprint arXiv:2411.09583},
  year={2024}
}

@article{Christiansen2021inverse,
  author  = {Christiansen, Rasmus E. and Lin, Zin and Roques-Carmes, Charles and Salamin, Yannick and Johnson, Steven G. and Solja{\v{c}}i{\'c}, Marin},
  title   = {Fullwave {M}axwell inverse design of axisymmetric, tunable, and multi-scale multi-wavelength metalenses},
  journal = {Optics Express},
  volume  = {28},
  number  = {23},
  pages   = {33854--33868},
  year    = {2020},
  doi     = {10.1364/OE.403192}
}

@book{Harrington1961,
  author    = {Harrington, Roger F.},
  title     = {Time-Harmonic Electromagnetic Fields},
  publisher = {McGraw-Hill},
  year      = {1961}
}

@article{Lalanne1998,
  author  = {Lalanne, Philippe and Astilean, Simion and Chavel, Pierre and Cambril, Edmond and Launois, Huguette},
  title   = {Design and fabrication of blazed binary diffractive elements with sampling periods smaller than the structural cutoff},
  journal = {Journal of the Optical Society of America A},
  volume  = {16},
  number  = {5},
  pages   = {1143--1156},
  year    = {1999},
  doi     = {10.1364/JOSAA.16.001143}
}

@article{PestourieSurrogate2022,
  author  = {Pestourie, Rapha{\"e}l and Mroueh, Youssef and Rackauckas, Chris and Das, Payel and Johnson, Steven G.},
  title   = {Physics-enhanced deep surrogates for partial differential equations},
  journal = {Nature Machine Intelligence},
  volume  = {5},
  pages   = {1458--1465},
  year    = {2023},
  doi     = {10.1038/s42256-023-00761-y}
}

@article{Mansouree2021,
  author  = {Mansouree, Mahdad and McClung, Andrew and Samudrala, Sarath and Arbabi, Amir},
  title   = {Large-Scale Parametrized Metasurface Design Using Adjoint Optimization},
  journal = {ACS Photonics},
  volume  = {8},
  number  = {2},
  pages   = {455--463},
  year    = {2021},
  doi     = {10.1021/acsphotonics.0c01058}
}

@article{iserles2004quadrature,
  title={On quadrature methods for highly oscillatory integrals and their implementation},
  author={Iserles, Arieh and N{\o}rsett, Syvert P},
  journal={BIT Numerical Mathematics},
  volume={44},
  number={4},
  pages={755--772},
  year={2004},
  publisher={Springer}
}

@article{Ma2021deep,
  author  = {Ma, Wei and Liu, Zhaocheng and Kudyshev, Zhaxylyk A. and Boltasseva, Alexandra and Cai, Wenshan and Liu, Yongmin},
  title   = {Deep learning for the design of photonic structures},
  journal = {Nature Photonics},
  volume  = {15},
  pages   = {77--90},
  year    = {2021},
  doi     = {10.1038/s41566-020-0685-y}
}

@article{So2020deep,
  author  = {So, Sunae and Badloe, Trevon and Noh, Junsuk and Bravo-Abad, Jorge and Rho, Junhwa},
  title   = {Deep learning enabled inverse design in nanophotonics},
  journal = {Nanophotonics},
  volume  = {9},
  number  = {5},
  pages   = {1041--1057},
  year    = {2020},
  doi     = {10.1515/nanoph-2019-0474}
}

@article{AnNeuralAdj2020,
  author  = {An, Sichao and Zheng, Bowen and Shalaginov, Mikhail Y. and Tang, Hong and Li, Hang and Zhou, Li and Ding, Jie and Agarwal, Anuradha M. and Rivero-Baleine, Clara and Kang, Myungkoo and Richardson, Kathleen A. and Gu, Tian and Hu, Juejun and Fowler, Clayton and Zhang, Hualiang},
  title   = {Deep learning modeling approach for metasurfaces with high degrees of freedom},
  journal = {Optics Express},
  volume  = {28},
  number  = {21},
  pages   = {31932--31942},
  year    = {2020},
  doi     = {10.1364/OE.401960}
}

@article{lin2021computational,
  title={Computational inverse design for ultra-compact single-piece metalenses free of chromatic and angular aberration},
  author={Lin, Zin and Roques-Carmes, Charles and Christiansen, Rasmus E and Solja{\v{c}}i{\'c}, Marin and Johnson, Steven G},
  journal={Applied Physics Letters},
  volume={118},
  number={4},
  year={2021},
  publisher={AIP Publishing}
}

@article{lin2022end,
  title={End-to-end metasurface inverse design for single-shot multi-channel imaging},
  author={Lin, Zin and Pestourie, Rapha{\"e}l and Roques-Carmes, Charles and Li, Zhaoyi and Capasso, Federico and Solja{\v{c}}i{\'c}, Marin and Johnson, Steven G},
  journal={Optics express},
  volume={30},
  number={16},
  pages={28358--28370},
  year={2022},
  publisher={Optica Publishing Group}
}

@article{lin2019overlapping,
  title={Overlapping domains for topology optimization of large-area metasurfaces},
  author={Lin, Zin and Johnson, Steven G},
  journal={Optics express},
  volume={27},
  number={22},
  pages={32445--32453},
  year={2019},
  publisher={Optical Society of America}
}

@article{sun2025scalable,
  author    = {Sun, Mengdi and Shakeri, Ata and Keshvari, Arvin and Giannakopoulos, Dimitrios and Wang, Qing and Chen, Wei-Ting and Johnson, Steven G. and Lin, Zin},
  title     = {Scalable Freeform Optimization of Wide-Aperture {3D} Metalenses by Zoned Discrete Axisymmetry},
  journal   = {ACS Photonics},
  volume    = {12},
  number    = {6},
  pages     = {3163--3171},
  year      = {2025},
  publisher = {ACS Publications},
  doi       = {10.1021/acsphotonics.5c00505}
}

@article{Isnard:25,
author = {Enzo Isnard and S\'{e}bastien H\'{e}ron and St\'{e}phane Lanteri and Mahmoud Elsawy},
journal = {Opt. Express},
number = {25},
pages = {52600--52613},
publisher = {Optica Publishing Group},
title = {Hybrid model to simulate optical systems combining metasurfaces and classical refractive elements},
volume = {33},
month = {Dec},
year = {2025},
url = {https://opg.optica.org/oe/abstract.cfm?URI=oe-33-25-52600},
doi = {10.1364/OE.580729}
}

@software{2015ascl.soft12017H,
       author = {{Hamilton}, Andrew J.~S.},
        title = "{FFTLog: Fast Fourier or Hankel transform}",
 howpublished = {Astrophysics Source Code Library, record ascl:1512.017},
         year = 2015,
        month = dec,
          eid = {ascl:1512.017},
archivePrefix = {ascl},
       eprint = {1512.017},
       adsurl = {https://ui.adsabs.harvard.edu/abs/2015ascl.soft12017H}
}

@article{lozier2003nist,
  title={NIST digital library of mathematical functions},
  author={Lozier, Daniel W},
  journal={Annals of Mathematics and Artificial Intelligence},
  volume={38},
  number={1},
  pages={105--119},
  year={2003},
  publisher={Springer}
}

@article{wei2024spatially,
  title={Spatially varying nanophotonic neural networks},
  author={Wei, Kaixuan and Li, Xiao and Froech, Johannes and Chakravarthula, Praneeth and Whitehead, James and Tseng, Ethan and Majumdar, Arka and Heide, Felix},
  journal={Science Advances},
  volume={10},
  number={45},
  pages={eadp0391},
  year={2024},
  publisher={American Association for the Advancement of Science}
}

@article{wirth2025wide,
  title={Wide field of view large aperture meta-doublet eyepiece},
  author={Wirth-Singh, Anna and Fr{\"o}ch, Johannes E and Yang, Fan and Martin, Louis and Zheng, Hanyu and Zhang, Hualiang and Tanguy, Quentin T and Zhou, Zhihao and Huang, Luocheng and John, Demis D and others},
  journal={Light: Science \& Applications},
  volume={14},
  number={1},
  pages={17},
  year={2025},
  publisher={Nature Publishing Group UK London}
}

@article{zerbe2026analytical,
  title={Analytical framework for angular dispersion engineering of metasurface optics},
  author={Zerbe, Benjamin A and Campbell, Sawyer D and Werner, Douglas H},
  journal={Optics Express},
  volume={34},
  number={5},
  pages={7967--7978},
  year={2026},
  publisher={Optica Publishing Group}
}

@article{Hamilton2000,
  author  = {Hamilton, A. J. S.},
  title   = {Uncorrelated modes of the non-linear power spectrum},
  journal = {Monthly Notices of the Royal Astronomical Society},
  volume  = {312},
  number  = {2},
  pages   = {257--284},
  year    = {2000},
  doi     = {10.1046/j.1365-8711.2000.03071.x}
}

@book{Watson1944,
  author    = {Watson, G. N.},
  title     = {A Treatise on the Theory of Bessel Functions},
  edition   = {2nd},
  publisher = {Cambridge University Press},
  year      = {1944}
}

\end{document}